\documentclass[12pt]{article}
\pdfoutput=1
\usepackage{jheppub}
\usepackage[utf8]{inputenc}
\usepackage{verbatim}
\usepackage{amsmath}
\usepackage{amssymb}
\usepackage{amsthm}
\usepackage{slashed}
\usepackage{amsfonts}

\newcommand{\be}{\begin{equation}}
\newcommand{\ee}{\end{equation}}
\newcommand{\bfig}{\begin{figure}\begin{center}}
\newcommand{\efig}{\end{center}\end{figure}}
\newcommand{\bi}{\begin{itemize}}
\newcommand{\ei}{\end{itemize}}

\newcommand{\wt}{\widetilde}

\newcommand{\Pc}{\mathcal{P}}

\theoremstyle{definition}

\begin{document}
\title{Covariant phase space with boundaries}
\author[a]{Daniel Harlow}
\author[a]{and Jie-qiang Wu}
\affiliation[a]{Center for Theoretical Physics\\ Massachusetts Institute of Technology, Cambridge, MA 02139, USA}
\emailAdd{harlow@mit.edu, jieqiang@mit.edu}
\abstract{The covariant phase space method of Iyer, Lee, Wald, and Zoupas gives an elegant way to understand the Hamiltonian dynamics of Lagrangian field theories without breaking covariance.  The original literature however does not systematically treat total derivatives and boundary terms, which has led to some confusion about how exactly to apply the formalism in the presence of boundaries.  In particular the original construction of the canonical Hamiltonian relies on the assumed existence of a certain boundary quantity ``$B$'', whose physical interpretation has not been clear.  We here give an algorithmic procedure for applying the covariant phase space formalism to field theories with spatial boundaries, from which the term in the Hamiltonian involving $B$ emerges naturally.  Our procedure also produces an additional boundary term, which was not present in the original literature and which so far has only appeared implicitly in specific examples, and which is already nonvanishing even in general relativity with sufficiently permissive boundary conditions. The only requirement we impose is that at solutions of the equations of motion the action is stationary modulo future/past boundary terms under arbitrary variations obeying the spatial boundary conditions; from this the symplectic structure and the Hamiltonian for any diffeomorphism  that preserves the theory are unambiguously constructed.  We show in examples that the Hamiltonian so constructed agrees with previous results.  We also show that the Poisson bracket on covariant phase space directly coincides with the Peierls bracket, without any need for non-covariant intermediate steps, and we discuss possible implications for the entropy of dynamical black hole horizons.}
\maketitle
\section{Introduction}
The most basic problem in physics is the initial-value problem: given the state of a system at some initial time, in what state do we find it at a later time?  This problem is most naturally discussed within the Hamiltonian formulation of classical/quantum mechanics.  In relativistic theories however it is difficult to use this formalism without destroying manifest covariance: any straightforward approach requires one to pick a preferred set of time slices.  Such a choice is especially inconvenient in theories which are generally-covariant, such as Einstein's theory of gravity.

The standard approach to this problem is to de-emphasize the Hamiltonian formalism, restricting classically to Lagrangians and quantum mechanically to path integrals.  This works fine for many applications, but there remain some topics, such as the initial-value problem, for which the Hamiltonian formalism is too convenient to dispense with.  For example it is only in the Hamiltonian formalism that one can do a proper accounting of the degrees of freedom in a system, and thermodynamic quantities such as energy and entropy are naturally defined there.

In relativistic field theories there is an elegant formalism due to Iyer, Lee, Wald, and Zoupas, which, building on earlier ideas from \cite{Witten:1986qs,zuckerman1987action,crnkovic1987covariant,Crnkovic:1987tz}, presents Hamiltonian mechanics in a manner that preserves manifest Lorentz or diffeomorphism invariance: the \textit{covariant phase space formalism} \cite{Lee:1990nz,Wald:1993nt,Iyer:1994ys,Iyer:1995kg,Wald:1999wa}.\footnote{This description of the history is somewhat over-simplified, see the introduction of \cite{Khavkine:2014kya} for a more detailed discussion of the antecedents of the formalism (which at least go back to ideas of Bergmann in the 1950s).  Also the construction of phase space in \cite{Witten:1986qs,zuckerman1987action,crnkovic1987covariant,Crnkovic:1987tz} proceeds in a more direct manner than that in \cite{Lee:1990nz,Wald:1993nt,Iyer:1994ys,Iyer:1995kg,Wald:1999wa}: the former first restricts to solutions of the equations of motion and then quotients by the zero-modes of a pre-symplectic form on those solutions, while the latter first quotients by zero modes of the pre-symplectic form on configuration space, then imposes the equations of motion, and then performs a further quotient by any new zero modes which appeared.  In this paper we will adopt the simpler first approach, but fortunately most equations are the same either way.}  This method is well-known in the relativity community, where in particular it was used by Wald to derive a generalization of the area formula for black hole entropy to higher-derivative gravity \cite{Wald:1993nt}, and it has been showing up fairly often in recent discussions of the AdS/CFT correspondence (see e.g. \cite{Hollands:2005wt,Compere:2008us,Faulkner:2013ica,Andrade:2015gja,Jafferis:2015del,Lashkari:2016idm,Dong:2018seb,Belin:2018fxe,Belin:2018bpg}), the asymptotic symmetry structure of gravity in Minkowski space
\cite{Compere:2011ve,Chandrasekaran:2018aop}, and in attempts to define ``near-horizon'' symmetries associated to black holes \cite{Carlip:1999cy,Haco:2018ske,Haco:2019ggi}.

This note grew out of the authors' attempts to understand the covariant phase space formalism.  Its primary goal is pedagogical: to present that formalism in a way that avoids some confusions which the authors, and apparently also others, ran into in studying the original literature.  These confusions have to do with the role of boundary terms and total derivatives in the formalism, which in the standard presentation \cite{Iyer:1994ys} were treated in a somewhat cavalier manner.  Indeed in \cite{Iyer:1994ys} boundary terms and total derivatives were ignored for most of the initial discussion, but then the existence of the Hamiltonian was presented as requiring the existence of a boundary quantity called $B$ obeying a certain integrability condition.\footnote{This was also the style of argument in the classic discussion \cite{Regge:1974zd} of the asymptotic symmetries of general relativity in asymptotically-flat space, where (using non-covariant techniques) the form of the Hamiltonian was motivated using consistency requirements instead of derived systematically.}  Moreover no general reassurance as to when such a quantity exists was given, which is surprising from the point of view of the ordinary canonical formalism: usually the Hamiltonian can be obtained from the Lagrangian algorithmically via the equation $H=p_a\dot{q}^a-L$.  In a formalism which treats boundary terms systematically, the existence of the Hamiltonian should be automatic (as for example is the case in the non-covariant analysis of general relativity given in \cite{Hawking:1995fd,Hawking:1996ww}).  Our goal in this note is to give such a systematic treatment within the covariant phase space formalism. As a bonus, we will find that the formula given in \cite{Iyer:1994ys} for the canonical Hamiltonian is not correct in general: there is an additional boundary term which is nonzero even in general relativity for sufficiently permissive boundary conditions, and which is generically nonzero for theories with sufficiently many derivatives.  After presenting our general formalism, we illustrate it in several examples, recovering known results.  

We emphasize that in this paper, the boundary conditions at any spatial boundaries are viewed as part of the definition of a field theory.  For example a scalar field in a cavity with Dirichlet walls and a scalar field in a cavity with Neumann walls are different Hamiltonian systems.  The Hamiltonian formulation of mechanics is global in nature, so to construct it properly we need to say what the rules are everywhere in space.    To avoid the question of convergence we have written most of the paper assuming that any spatial boundaries are finitely far away.  Finite boundaries are of direct physical relevance e.g. in condensed matter systems and electromagnetic cavities, and they are also sensible in the context of linearized gravity.  On the other hand finite boundaries are difficult to implement in non-linear gravity (what would happen when a black hole meets a finite boundary?), and it is more natural to consider ``asymptotic'' boundaries that are infinitely far away. The logic of our paper should apply to asymptotic boundaries as well provided one is careful about manipulating infinite quantities; we discuss this further in section \ref{asympsec} at the end of the paper.

Our results are simple enough that we can briefly describe them here.  Indeed we consider a classical field theory action
\be
S=\int_M L+\int_{\partial M} \ell,
\ee
where $L$ is a $d$-form and $\ell$ is a $(d-1)$-form.  $\partial M$ in general includes both spatial and future/past pieces, in this paper we do not consider null boundaries.  The variation of $L$ always has the form
\be
\delta L =E_a\delta\phi^a+d\Theta,
\ee
where $E_a=0$ are the equations of motion and $\Theta$ is a $(d-1)$-form which is linear in the variations of the dynamical fields $\phi^a$.  Stationarity of the action up to future/past boundary terms requires
\be
\left(\Theta+\delta \ell\right)|_{\Gamma}=dC,
\ee
where $\Gamma$ is the spatial boundary and $C$ is a $(d-2)$-form defined on $\Gamma$ that is also linear in the field variations.  The (pre-)symplectic form of this system is given by
\be
\wt{\Omega}=\int_\Sigma\delta\left(\Theta-d C\right),
\ee
where $\Sigma$ is a Cauchy slice and the precise meaning of the second variation implicit in this formula is explained below (basically we re-interpret $\delta$ as the exterior derivative on the space of field configurations).  Finally if $\xi^\mu$ is a vector field generating a one-parameter family of diffeomorphisms which preserve the boundary conditions, and under which $L$, $\ell$, and $C$ transform covariantly, then the Hamiltonian which generates this family of diffeomorphisms is given by
\be
H_\xi=\int_\Sigma J_\xi +\int_{\partial \Sigma}\left(\xi \cdot \ell-X_\xi \cdot C\right).
\ee
Here ``$\xi\cdot\ell$'' indicates insertion of $\xi$ into the first argument of $\ell$, ``$X_\xi \cdot C$'' denotes replacing $\delta\phi^a$ in $C$ by the Lie derivative $\mathcal{L}_\xi \phi^a$, and $J_\xi=X_\xi\cdot \Theta-\xi\cdot L$ is the ``Noether current''.  In theories where $L$ is covariant under arbitrary diffeomorphisms, such as general relativity, it was shown in \cite{wald1990identically,Iyer:1994ys} that there must be a local $(d-2)$-form $Q_\xi$ such that $J_\xi=dQ_\xi$. Thus in such theories the Hamiltonian conjugate to $\xi$ is a pure boundary term:
\be
H_\xi=\int_{\partial \Sigma}\left(Q_\xi+\xi \cdot \ell-X_\xi \cdot C\right).
\ee
The remainder of this paper explains these formulas in more detail and illustrates them using examples.  In a final section we show that the Poisson bracket in the covariant phase space formalism is generally equivalent to the Peierls bracket, we give a proof of Noether's theorem for continuous symmetries within the covariant phase space approach,  and we comment on some subtleties arising in the application of our results to asymptotic boundaries.

The inclusion of boundary terms in the covariant phase space formalism was previously considered in \cite{Iyer:1995kg,Julia:2002df,Compere:2008us,Papadimitriou:2005ii,Compere:2011ve,Andrade:2015fna,Andrade:2015gja,Donnelly:2016rvo,Giddings:2018umg}, each of which has some nontrivial overlap with our discussion.  In particular setting $C=0$ in our formalism one obtains a formalism described in \cite{Iyer:1995kg}, but as we explain below this is an inappropriate restriction.  A formalism with nonzero $C$ was introduced in \cite{Compere:2008us,Compere:2011ve,Andrade:2015gja}, but the covariance properties of $C$ were not studied and its contribution to canonical charges such as the Hamiltonian was shown only in general relativity with specific boundary conditions.  An alternative formalism in which many of the same issues can be addressed was given in \cite{Barnich:2001jy,Barnich:2007bf}; we have not studied in detail the relationship between that formalism and ours, but it requires integrability assumptions of the type we avoid and the treatment of boundary terms seems to be less general than ours.\footnote{We thank Geoffrey Comp\`{e}re for explaining several aspects of the formalism of \cite{Barnich:2001jy,Barnich:2007bf}.} Effects which can be interpreted as arising from our $C$ term were found for general relativity with a noncompact asymptotic boundary in \cite{almaraz2014positive,almaraz2018mass,almaraz2019spacetime}.  We believe our treatment of boundary terms is the most complete so far, and also perhaps the most efficient.  We have not systematically treated fermionic fields or topologically nontrivial gauge connections, but we foresee no difficulty with incorporating them along the lines of \cite{Prabhu:2015vua}.

\subsection{Notation}\label{notation}
In this paper we make heavy use of differential forms, our conventions for these are that if $\omega$ is a $p$ form and $\sigma$ is a $q$ form, we have
\begin{align}\nonumber
\left(\omega\wedge\sigma\right)_{\mu_1\ldots \mu_p\nu_1\ldots \nu_q}&=\frac{(p+q)!}{p!q!}\omega_{[\mu_1\ldots \mu_p}\sigma_{\nu_1\ldots \nu_q]}\\\nonumber
(d\omega)_{\mu_0\ldots \mu_p}&=(p+1)\partial_{[\mu_0}\omega_{\mu_1\ldots \mu_p]}\\
(\star\omega)_{\mu_1\ldots \mu_{d-p}}&=\frac{1}{p!}\epsilon^{\nu_1\ldots \nu_p}_{\phantom{\nu_1\ldots \nu_p}\mu_1\ldots \mu_{d-p}}\omega_{\nu_1\ldots \nu_p}.
\end{align}
Here ``$[\cdot]''$ denotes averaging over index permutations weighted by sign, so for example $\omega_{[\mu}\sigma_{\nu]}=\frac{1}{2}\left(\omega_\mu\sigma_\nu-\omega_\nu\sigma_\mu\right)$, and $\epsilon$ is the volume form.  The Lie derivative of any differential form $\omega$ with respect to a vector field $X$ is related to the exterior derivative via Cartan's magic formula
\be\label{cartan}
\mathcal{L}_X \omega=X\cdot d\omega+d(X\cdot \omega),
\ee
where $\cdot$ denotes inserting a vector into the first argument of a differential form (if $\omega$ is a zero-form we define $X\cdot \omega=0$).  Throughout the paper we will use ``$d$'' to indicate the exterior derivative on spacetime and ``$\delta$'' to indicate the exterior derivative on configuration space (and also its pullback to pre-phase space and phase space), a notation we discuss further around equation \eqref{deltaint}.

We take spacetime to be a manifold with boundary $M$, whose boundary we call $\partial M$, and we are often interested a Cauchy surface $\Sigma$ and its boundary $\partial \Sigma$.  We here set up some conventions about how to assign orientations to these various submanifolds of $M$.  Given an orientation on an orientable manifold with boundary $M$, there is a natural orientation induced on $\partial M$ such that Stokes' theorem
\be\label{stokes}
\int_M d\omega=\int_{\partial M} \omega
\ee
holds.  If $M$ has a metric, as it always will for us, then we can describe this induced orientation by saying we require that the boundary volume form $\epsilon_{\partial M}$ is related to the spacetime volume form $\epsilon$ by
\be
\epsilon=n\wedge \epsilon_{\partial M}, \label{epsilonrelation}
\ee
where $n$ is the ``outward pointing'' normal form defined by equation \eqref{normaldef} below.  We will always use this orientation for $\partial M$. We will also adopt the orientation on $\Sigma$ given by viewing it as the boundary of its past in $M$, and we will adopt the orientation on $\partial \Sigma$ given by viewing it as the boundary of $\Sigma$.   So for example if we take $M$ to be the region with $x\leq 0$ in Minkowski space, with volume form $\epsilon=dt\wedge dx \wedge dy \wedge dz$, and we take $\Sigma$ to be the surface $t=0$, then the volume form $\epsilon_{\partial M}$ on $\partial M$ is $-dt\wedge dy \wedge dz$, the volume form $\epsilon_\Sigma$ on $\Sigma$ is $dx\wedge dy \wedge dz$, and the volume form $\epsilon_{\partial \Sigma}$ on $\partial \Sigma$ is $dy\wedge dz$.  Note in particular that the volume form on $\partial \Sigma$ is \textit{not} obtained by viewing $\partial \Sigma$ as the boundary of its past within $\partial M$, these differ by a sign.  Sometimes we will discuss a Cauchy surface $\Sigma_-$ which is the past boundary of a spacetime  $M$, the most convenient way to maintain our conventions is to say that when this surface appears implicitly as part of $\partial M$ we give it the opposite orientation from when it appears explicitly as $\Sigma_-$.

\section{Formalism}
\subsection{Hamiltonian mechanics}
Hamiltonian mechanics is often presented as the dynamics of a phase space labeled by position and momentum coordinates $q^a$, $p^a$, with any scalar function $H$ on this phase space generating dynamical evolution via Hamilton's equations
\begin{align}\nonumber
\dot{q}^a&=\frac{\partial H}{\partial p^a}\\
\dot{p}^a&=-\frac{\partial H}{\partial q^a}.
\end{align}
Unfortunately this split of coordinates into positions and momenta makes it difficult to preserve covariance.  There is however an elegant geometric formulation of Hamiltonian mechanics  which allows us to avoid making such a split.  Namely we instead view phase space as an abstract manifold $\Pc$, endowed with a closed non-degenerate two-form $\Omega$ called the \textit{symplectic form} \cite{arnold2007mathematical}.  A manifold equipped with such a form is called a \textit{symplectic manifold}.  We now briefly review Hamiltonian mechanics from this point of view.

Let $\Pc$ be a symplectic manifold, with symplectic form $\Omega$.  We can view $\Omega$ as a map from vectors to one-forms via $\Omega(Y)(X)\equiv\Omega(X,Y)$, and since $\Omega$ is non-degenerate this map will have an inverse, $\Omega^{-1}$, which we can also view as an anti-symmetric two-vector mapping a pair of one-forms to a real number via $\Omega^{-1}(\omega,\sigma)\equiv\omega(\Omega^{-1}(\sigma))$.  Given any function $H:\Pc\to \mathbb{R}$, we can then define a vector field $X_H$ on $\Pc$ via
\be\label{XHeq}
X_H(f)\equiv\Omega^{-1}(\delta f,\delta H),
\ee
where $f:\Pc\to\mathbb{R}$ is an arbitrary function on $\Pc$.
Here we introduce a notation where we denote the exterior derivative on phase space by $\delta$ to distinguish it from the exterior derivative $d$ on spacetime which appears below.  The idea is then to view the integral curves of $X_H$ in $\Pc$ as giving the time evolution of the system generated by viewing $H$ as the Hamiltonian.  We can express this using the \textit{Poisson bracket} of two functions $f$ and $g$ on $\Pc$, defined by
\be
\{f,g\}\equiv \Omega^{-1}(\delta f,\delta g)=\Omega(X_g,X_f),
\ee
in terms of which we have the time evolution
\be
\dot{f}\equiv X_H(f)=\{f,H\}
\ee
for any function $f:\Pc\mapsto \mathbb{R}$.  Clearly $\Omega$ must be non-degenerate for this dynamics to be well-defined.  It is less obvious why $\Omega$ is required to be closed, and in fact there are dynamical systems where it isn't, but in such systems the Poisson bracket is not preserved under time evolution by an arbitrary Hamiltonian so it cannot become a commutator in quantum mechanics.\footnote{One way to see this is the following: conservation of the Poisson bracket is equivalent to saying that the Lie derivative $\mathcal{L}_{X_H}\Omega$ vanishes.    From \eqref{cartan} we then have
\begin{align}\nonumber
\mathcal{L}_{X_H}\Omega&=X_H\cdot \delta \Omega+\delta(X_H\cdot \Omega)\\\nonumber
&=X_H\cdot \delta \Omega+\delta(-\delta H)\\
&=X_H\cdot \delta \Omega,
\end{align}
so for this to vanish for arbitrary $H$ we need $\delta \Omega=0$.}
The old-fashioned version of the Hamiltonian formalism using $q^a$ and $p^a$ is recovered from these definitions by taking
\be\label{pqomega}
\Omega=\sum_a \delta p^a\wedge \delta q^a.
\ee

The standard interpretation of the phase space of a dynamical system is that it labels the set of distinct initial conditions on a time slice.  This interpretation is not covariant, as we need to specify the time slice.  The main idea of the covariant phase space formalism, going back (at least) to \cite{Witten:1986qs,zuckerman1987action,crnkovic1987covariant,Crnkovic:1987tz}, is, roughly speaking, to instead define phase space as the set of solutions of the equations of motion.  To the extent that the initial value problem is well-defined, these should be in one-to-one correspondence with the set of initial conditions on any time slice.  This definition however needs some improvement for theories with continuous local symmetries, since for such theories the initial value problem is not well-defined \cite{Noether:1918zz}.\footnote{Continuous local symmetries disrupt the ``uniqueness'' part of the initial value problem; one also has to worry about the ``existence'' part.  We discuss this further at the end of section \ref{chargesec}.}  For example a solution $A_\mu$ of Maxwell's equations can always be turned into another equally good solution by a gauge transformation which has zero support in a neighborhood of any particular time slice.  In the above language, this problem arises because the naive symplectic form one derives from the Maxwell Lagrangian is degenerate (we review this example further in section \ref{maxwellsec} below).

Fortunately there is a nice way to deal with this: one instead refers to the set of solutions of the equations of motion (obeying any needed boundary conditions) as \textit{pre-phase space} $\wt{\Pc}$, and the phase space $\Pc$ is then obtained by an appropriate quotient.  We will soon see that in any Lagrangian field theory this pre-phase space is always naturally equipped with a \textit{pre-symplectic form} $\wt{\Omega}$, which is a closed two-form on $\wt{\Pc}$ that we will assume has constant but not necessarily full rank.  The physical phase space $\Pc$ is then obtained by quotienting $\wt{\Pc}$ by the action of the group of continuous transformations whose generators are zero modes of $\wt{\Omega}$ \cite{Witten:1986qs,zuckerman1987action,crnkovic1987covariant,Crnkovic:1987tz}.\footnote{We again highlight here the alternative covariant phase space construction used in \cite{Lee:1990nz}, where prior to imposing the equations of motion one already quotients the set of all field configurations by the zero modes of the pre-symplectic form.  One then has to perform a second quotient over any new zero modes which appear after the equations of motion are imposed. This approach can be useful in identifying the set of valid initial data on a Cauchy slice, but we have chosen to adopt the more direct construction of \cite{Witten:1986qs,zuckerman1987action,crnkovic1987covariant,Crnkovic:1987tz}.}  More explicitly, if $\wt{X}$ and $\wt{Y}$ are vector fields on $\wt{\Pc}$ which are everywhere annihilated by $\wt{\Omega}$, then their commutator $[\wt{X},\wt{Y}]\equiv \mathcal{L}_{\wt{X}} \wt{Y}$ will also be everywhere annihilated by $\wt{\Omega}$.  Indeed using $\delta\wt{\Omega}=0$, $\wt{X}\cdot \wt{\Omega}=\wt{Y}\cdot \wt{\Omega}=0$, and \eqref{cartan}, we have
\begin{align}\nonumber
\mathcal{L}_{\wt{X}} \wt{Y}\cdot \wt{\Omega}&=\mathcal{L}_{\wt{X}}(\wt{Y}\cdot \wt{\Omega})-\wt{Y}\cdot \mathcal{L}_{\wt{X}}\wt{\Omega}\\\nonumber
&=-\wt{Y}\cdot \left(\wt{X}\cdot \delta\wt{\Omega}+\delta(\wt{X}\cdot \wt{\Omega})\right)\\
&=0.
\end{align}
The set of zero-mode vector fields of $\wt{\Omega}$ thus form a (possibly infinite-dimensional) Lie algebra, and by Frobenius's theorem they are jointly tangent to a set of submanifolds which foliate $\wt{\Pc}$.  These submanifolds can be thought of as the orbits of the connected subgroup $\wt{G}$ of the diffeomorphisms of $\wt{\Pc}$ whose Lie algebra corresponds to the zero modes of $\wt{\Omega}$.  The physical phase space $\Pc$ is then defined as the quotient of $\wt{\Pc}$ by this action:
\be
\Pc\equiv \wt{\Pc}/\wt{G}.
\ee
Thus the action of $\wt{G}$ is a redundancy of description that leaves no imprint on $\Pc$; in local field theories it is typically realized as a set of continuous gauge transformations which become trivial sufficiently quickly at any boundaries.\footnote{Discrete gauge symmetries do not lead to zero modes of the pre-symplectic form, but in going from $\wt{\Pc}$ to $\Pc$ we should still quotient by some or all of them depending on the boundary conditions.}

To complete the construction of the phase space $\Pc$, we must also define a symplectic form $\Omega$.   This is done in the following way.   Let $\pi:\wt{\Pc}\to\Pc$ be the map that that sends each point in $\wt{\Pc}$ to its $\wt{G}$-orbit, let $p$ be a point in $\Pc$, and let $X$ and $Y$ be vectors in the tangent space $T_p \Pc$.  We can always find a point $q\in \wt{\Pc}$ and vectors $\wt{X}$ and $\wt{Y}$ in $T_q \wt{\Pc}$ such that $X$ and $Y$ are the pushforwards of $\wt{X}$ and $\wt{Y}$ by $\pi$.  We then define
\be\label{Omegadefgen}
\Omega(X,Y)\equiv \wt{\Omega}(\wt{X},\wt{Y}).
\ee
For this $\Omega$ to be well-defined, we need to show that it is independent of the arbitrariness involved in choosing $q$, $\wt{X}$, and $\wt{Y}$.  We first note that two vectors $\wt{X}$ and $\wt{X}'$ in $T_q\wt{\Pc}$ which both push forward to the same $X\in T_p \Pc$ can differ only by addition of a vector annihilated by $\wt{\Omega}$ at $p$: this ambiguity thus has no effect in \eqref{Omegadefgen}.  Secondly we observe that by definition any two points $q, q'\in \wt{\Pc}$ which both map to $p$ are related by the group action: $q'=gq$ for some $g\in \wt{G}$.  This implies that the pushforward $\wt{X}'\in T_{q'} \wt{\Pc}$ of $\wt{X}\in T_q\wt{\Pc}$ by $g$  maps via pushforward by $\pi$ to the same element of $T_p \Pc$ that $\wt{X}$ does (this follows from $\pi\circ g=\pi$). Moreover $\wt{\Omega}$ is invariant under pushforward by $g$: this follows from the fact that by \eqref{cartan} for any zero mode $\wt{X}$ of $\wt{\Omega}$ we have
\be
\mathcal{L}_{\wt{X}}\wt{\Omega}=\wt{X}\cdot \delta\wt{\Omega}+\delta(\wt{X}\cdot \wt{\Omega})=0.
\ee
Together these results imply that \eqref{Omegadefgen} is indeed unambiguous.  Finally we argue that $\Omega$ is non-degenerate.  Indeed let's assume that for some $p\in \Pc$ there exists $X\in T_p \Pc$ such that $X\neq 0$ but $X\cdot \Omega=0$.  This $X$ must be the pushforward of some $\wt{X}\in T_q\wt{\Pc}$, and by \eqref{Omegadefgen} we must have $\wt{X}\cdot \wt{\Omega}=0$.  Since $\wt{\Omega}$ has constant rank, we can extend $\wt{X}$ to a vector field which is annihilated by $\wt{\Omega}$ throughout $\Pc$.  Therefore the pushforward of this vector field by $\pi$ must vanish, which contradicts our assumption that $X\neq 0$. Thus $\Omega$ has full rank at each point in $\Pc$ and is indeed a symplectic form.

This discussion has so far been abstract; an example may be helpful.  Consider a free non-relativistic particle, with action
\be\label{partex}
S=\frac{m}{2}\int dt \dot{x}^2.
\ee
There is a two-parameter set of solutions
\be\label{partsol}
x(t)=\frac{p_0}{m}t+x_0,
\ee
so we can use $(x_0,p_0)$ as coordinates on phase space.  The symplectic form (here $\wt{G}$ is trivial so no quotient is needed) is $\delta p_0\wedge \delta x_0$, and the Hamiltonian evolution on this set of solutions generated by the Hamiltonian $H=\frac{p_0^2}{2m}$ is
\begin{align}\nonumber
p_0(t')&=p_0\\
x_0(t')&=\frac{p_0}{m}t'+x_0.\label{partev}
\end{align}
We emphasize the difference in interpretation between equations \eqref{partsol} and \eqref{partev}: the former gives a parametrization of the set of solutions by saying what is going on at $t=0$, while the latter gives an evolution on that set which is nontrivial even though each solution ``already knows'' its own evolution.\footnote{It may seem that the time $t=0$ is special here, but we only used it to choose coordinates on $\Pc$.  The evolution is defined geometrically, and can be described using whatever coordinates we like.} This distinction is especially clear if we evolve in this phase space using a Hamiltonian other than $\frac{p_0^2}{2m}$; we discuss this further in section \ref{discussion} below.

There are several mathematical subtleties in the construction of $\Pc$ and $\Omega$ which we will mention here but not address in detail.  First of all the manifolds $\wt{\Pc}$ and $\Pc$ are often infinite-dimensional, and thus require some care to properly manipulate.  We expect that a rigorous treatment based on interpreting them as Banach manifolds should be possible, as on such manifolds the main tools we use (exterior derivatives, Cartan's magic formula, vector flows, and Frobenius's theorem) continue to make sense \cite{lang2012fundamentals}, but we have not pursued this in detail so our treatment of these spaces should be viewed as heuristic.  Secondly there may be special points in $\wt{\Pc}$ which are invariant under a nontrivial subgroup of $\wt{G}$, in which case $\Pc$ will be singular at those configurations \cite{Crnkovic:1987tz,marsden1981lectures,abraham1978foundations}. For example in general relativity there can be special geometries which have continuous isometries, and if those isometries vanish in a neighborhood of any boundaries then they will correspond to zero modes of $\wt{\Omega}$ (this situation also violates our assumption of $\wt{\Omega}$ having constant rank, but we may also want to include discrete isometries in $\wt{G}$, for which fixed points do not imply a change in rank).  At worst however this affects only a measure zero set of points in $\Pc$, and even that seems unlikely to happen in asymptotically-AdS or asymptotically-flat spacetimes since isometries which are non-vanishing at the boundary are not generated by zero modes of $\wt{\Omega}$.  Finally $\wt{G}$ might fail to be a group due to flows which reach infinity in $\wt{P}$ in finite time (as might happen for solutions which develop singularities in finite time).\footnote{We thank Anton Kapustin for suggesting this possibility.}  In that case we can still define $\Pc$ as the set of submanifolds which are tangent to the zero modes of $\wt{\Omega}$, but its structure (and that of $\Omega$) may become more intricate.  We expect that the formalism could be sharpened to systematically address these issues, but in this paper we will not attempt it.\footnote{We emphasize however that our construction below of a pre-symplectic form $\wt{\Omega}$ and Hamiltonian $H_\xi$ obeying Hamilton's equation \eqref{inverseH} is rigorous: these mathematical issues are only potentially relevant once we try to interpret \eqref{inverseH} in terms of the theory of Hamiltonian flows on symplectic manifolds, which likely needs to be somewhat generalized to include all interesting examples.}

\subsection{Local Lagrangians}\label{locallagsec}
In Lagrangian field theories we can make the discussion of the previous section more concrete using the formalism of \cite{Lee:1990nz,Wald:1993nt,Iyer:1994ys,Iyer:1995kg,Wald:1999wa} (re-interpreted using the phase space construction of \cite{Witten:1986qs,zuckerman1987action,crnkovic1987covariant,Crnkovic:1987tz}).  In this formalism the Lagrangian density is converted into a Lagrangian $d$-form $L$,  which is a local functional of the dynamical fields $\phi$ and their derivatives, and also potentially of some non-dynamical background fields $\chi$ and their derivatives. For example for a self-interacting scalar field theory we have
\be\label{scalarL}
L=-\left(\frac{1}{2}\nabla_\mu\phi\nabla_\nu\phi g^{\mu\nu}+V(\phi)\right)\epsilon,
\ee
where $\phi$ is a dynamical field, $g_{\mu\nu}$ is a non-dynamical background metric, and $\epsilon$ is the spacetime volume form. To avoid confusion, we emphasize that in saying that $L$ is a $d$-form, we mean that it transforms as a $d$-form under diffeomorphisms which act on both the dynamical and background fields.  In the following subsection we will discuss the special case of \textit{covariant} Lagrangians, which transform as $d$-forms also under diffeomorphisms which act only the dynamical fields.

In \cite{Lee:1990nz,Wald:1993nt,Iyer:1994ys,Iyer:1995kg,Wald:1999wa} the Lagrangian form was viewed as only being defined up to the addition of a total derivative, but since we are being careful about boundary terms we will \textit{not} allow the Lagrangian to be arbitrarily modified by the addition of a total derivative.  Indeed when we integrate the Lagrangian $d$-form to define an action, we will include a boundary term obtained by integrating over $\partial M$ a $(d-1)$-form $\ell$ built out of the restrictions of $\phi$ and $\chi$ to the boundary $\partial M$, and also possibly their normal derivatives there:
\be\label{Sdef}
S=\int_M L+\int_{\partial M} \ell.
\ee
Thus we may shift $L$ by a total derivative only if we shift $\ell$ in a compensating manner that preserves $S$, and for the most part we will not do this.

The basic idea of Lagrangian mechanics is that, after imposing appropriate boundary conditions at $\partial M$, we should look for configurations $\phi_c$ about which the action is stationary under arbitrary variations of the dynamical fields which obey those boundary conditions.  In fact the truth is slightly more subtle, due to the fundamentally different meaning of boundary conditions at spatial boundaries and boundary conditions at future/past boundaries.  The former are part of the definition of the theory, while the latter specify a state within that theory.  If we wish to allow variations that change the state, which indeed we do, then we do not wish to impose any boundary conditions at future/past boundaries.  Stationarity of the action under such variations would be too strong of a requirement, typically it would lead to a problem with few or no solutions.  The right approach is instead to only require that the action be stationary up to terms which are localized at the future and past boundaries.  If we decompose $\partial M=\Gamma\cup \Sigma_-\cup \Sigma_+$, where $\Gamma$ is the spatial boundary, $\Sigma_-$ is the past boundary, and $\Sigma_+$ is the future boundary, then we should look for configurations $\Phi_c$ about which
\be\label{Svarreq}
\delta S=\int_{\Sigma_+}\Psi-\int_{\Sigma_-}\Psi,
\ee
where the variation obeys the boundary conditions at $\Gamma$ and $\Psi$ is locally constructed out of the dynamical and background fields at $\Sigma_{\pm}$.  In what follows it will be convenient to refer to the set of dynamical field configurations on spacetime obeying the boundary conditions at $\Gamma$, but not necessarily the equations of motion, as \textit{configuration space}, denoted by $\mathcal{C}$.  In classical mechanics configuration space is the arena in which the variational principle operates, while quantum mechanically it is the set of configurations one integrates over in the path integral.\footnote{In particle mechanics the term ``configuration space'' is sometimes used to describe the set of positions of particles at a fixed time.  Our configuration space instead is the set of possible histories for those particles prior to imposing the equations of motion.}

To discuss this more explicitly, it is convenient to note that, by way of ``integration by parts''-style manipulations, any local Lagrangian form must obey
\be\label{Lvar}
\delta L=E_a\delta \phi^a+d\Theta,
\ee
where $a$ is an index running over the dynamical fields $\phi^a$ (we are using the Einstein summation convention), $\delta\phi^a$ are variations of those fields in configuration space $\mathcal{C}$, $d$ is the spacetime exterior derivative, $\Theta$ is a local functional of the dynamical/background fields and their derivatives, and is also a homogeneous linear functional of the $\delta\phi^a$ and their derivatives.    The $E_a$ are local functionals of the dynamical/background fields and their derivatives.  $\Theta$ is a $(d-1)$-form on spacetime, and is called the \textit{symplectic potential}. It is defined only up to addition of a total derivative $dY$ for $Y$ some local $(d-2)$ form.  The variation of the action \eqref{Sdef} is thus
\be\label{Svar}
\delta S=\int_ME_a\delta \phi^a+\int_{\partial M} \left(\delta \ell+\Theta\right),
\ee
where we have used Stokes' theorem \eqref{stokes}.  For this to obey \eqref{Svarreq} for arbitrary variations obeying the boundary conditions at $\Gamma$ about a configuration $\phi_c$,  and since we can always adjust such variations arbitrarily in the interior of $M$, we see that $\phi_c$ must obey the \textit{equations of motion}
\be\label{EOM}
E_a[\phi_c]=0.
\ee
We moreover see that to avoid a term at the spatial boundary $\Gamma$ in \eqref{Svarreq}, we need the second term in \eqref{Svar} to only have support on $\Sigma_{\pm}$.  A first guess is that we therefore should require $(\delta \ell+\Theta)|_{\Gamma}=0$ for all  variations obeying the boundary conditions at $\Gamma$.  Given the ambiguity of shifting $\Theta$ by a total derivative, however, this is unnatural.  A more general sufficient condition, which we believe (but have not shown) is also necessary, is to require that
\be\label{Ceq}
(\Theta+\delta \ell)|_{\Gamma}=dC,
\ee
where $C$ is a local $(d-2)$-form on $\Gamma$ which is constructed from the dynamical/background fields, the variations $\delta\phi^a$, and derivatives of both.  As with $L$ and $\ell$, any addition to $\Theta$ of a total derivative $dY$ must be complemented by an addition of $Y$ to $C$, such that \eqref{Ceq} is preserved.\footnote{\label{redef}  Allowing $C\neq 0$ may at first seem like a trivial generalization, since after all we could extend $C$ arbitrarily into the interior of $M$ and then define $\Theta'=\Theta-dC$ and $C'=0$.  It will however be quite convenient below to take $\Theta$ to be covariant,  and $dC$ generally will not extend to $M$ in a covariant manner.  This is also the reason why we have not redefined $L'=L+d\ell$ and $\ell'=0$ to get rid of $\ell$ in the action.}  Making use of \eqref{Ceq} in \eqref{Svar}, we thus have 
\begin{align}\nonumber
\delta S&=\int_M E_a \delta \phi^a+\int_{\Sigma_+-\Sigma_-}\left(\Theta+\delta \ell\right)+\int_\Gamma \left(\Theta+\delta \ell\right)\\\nonumber
&=\int_M E_a \delta \phi^a+\int_{\Sigma_+-\Sigma_-}\left(\Theta+\delta \ell\right)+\int_{\partial \Gamma}C\\
&=\int_M E_a \delta \phi^a+\int_{\Sigma_+-\Sigma_-}\left(\Theta+\delta \ell-dC\right),
\end{align}
which is indeed of the form \eqref{Svarreq} for variations about configurations obeying the equations of motion $E_a=0$ with $\Psi=\Theta+\delta \ell-dC$ (writing it this way requires us to extend $C$ to $\Sigma_{\pm}$ in an arbitrary manner, but only its values at $\partial \Sigma_{\pm}$ actually contribute).

To set up the Hamiltonian formalism, we must now introduce a pre-phase space and pre-symplectic form.  We define the pre-phase space $\wt{\Pc}$ to be those elements of the configuration space $\mathcal{C}$ which also obey the equations of motion \eqref{EOM}.  We do not impose boundary conditions in the future/past, and any background field configurations are held fixed.  In defining the symplectic form, it is very useful to first note that there is a convenient change of notation which allows us to re-interpret quantities like $\Theta$ and $C$ as one-forms on $\mathcal{C}$ \cite{crnkovic1987covariant}.  The idea is that instead of viewing the quantity $\delta\phi^a(x)$ as an infinitesimal variation, we can (and from now on will) view it as a coordinate differential on $\mathcal{C}$.  In other words, $\delta$ now denotes the exterior derivative for differential forms on $\mathcal{C}$, and the action of $\delta\phi^a(x)$ on a vector field is given by
\be\label{deltaint}
\delta\phi^a(x) \left(\int d^dx'f^b(\phi,x')\frac{\delta}{\delta\phi^b(x')}\right)=f^a(\phi,x).
\ee
Thus if we wish to convert $\delta\phi^a(x)$ from a one-form back to a variation, we act with it on a vector whose components are the desired variation.  With this notation  $\Theta$ and $C$ are one-forms on configuration space, and we may then pull them back to one-forms on pre-phase space by restricting their action to those vectors which are tangent to $\wt{\Pc}$.\footnote{Tangent vectors to $\wt{\Pc}$ are precisely those whose components $f^a(\phi,x)$ obey the linearized equations of motion, in the sense that $E^a[\phi_c+f]=0$ to linear order in $f$.}

Using this new interpretation of $\delta$ we can now introduce our version of the \textit{pre-symplectic current} from \cite{Lee:1990nz,Wald:1993nt,Iyer:1994ys,Iyer:1995kg}, which we define as the pullback to $\wt{\Pc}$ of the quantity $\delta\Psi$:
\be\label{omegadef}
\omega\equiv\delta\Psi|_{\wt{\Pc}}=\delta(\Theta-dC)|_{\wt{\Pc}}.
\ee
Here we have used $\delta^2=0$.  Since the pullback and exterior derivative are commuting operations, $\omega$ is closed as a two-form on $\wt{\Pc}$.  Moreover $\omega$ vanishes on $\Gamma$, since by \eqref{Ceq} we have
\be\label{omegaboundary}
\omega|_{\Gamma}=\delta(\Theta+\delta \ell-dC)|_{\wt{\Pc},\Gamma}=0.
\ee
$\omega$ is also closed as a $(d-1)$-form on spacetime:
\be\label{omegaclosed}
d\omega=d\delta(\Theta-dC)=\delta d\Theta=\delta(\delta L-E_a\delta\phi^a)=-\delta E_a \wedge\delta \phi^a=0.
\ee
Here we have used that $E_a=0$ on $\wt{\Pc}$, and also that $d$ and $\delta$ commute.  Finally we define the \textit{pre-symplectic form} on $\wt{\Pc}$ as
\be\label{Omegadef}
\wt{\Omega}\equiv\int_{\Sigma}\omega,
\ee
where $\Sigma$ is any Cauchy slice of $M$.  \eqref{omegaboundary} and \eqref{omegaclosed} ensure that $\wt{\Omega}$ is independent of the choice of $\Sigma$.  Moreover from \eqref{omegadef} we have
\be
\wt{\Omega}=\delta\left(\int_\Sigma \Theta-\int_{\partial \Sigma} C\right),
\ee
so $\wt{\Omega}$ is independent of how we chose to extend $C$ into the interior of $M$. $\wt{\Omega}$ will be degenerate if there are continuous local symmetries, but once we quotient $\wt{\Pc}$ by the subgroup $\wt{G}$ of pre-phase space diffeomorphisms generated by the zero modes of $\wt{\Omega}$ (and possibly its extension by other discrete gauge symmetries) then the resulting symplectic form $\Omega$ on phase space will be non-degenerate (and closed since $\omega$ is closed on $\wt{\Pc}$).

\subsection{Covariant Lagrangians}\label{covlagsec}
The covariant phase space formalism is especially useful for systems whose dynamics are invariant under at least some continuous subgroup of the spacetime diffeomorphism group.  We first recall that by definition the variation of any dynamical tensor field $\phi$ under the infinitesimal diffeomorphism generated by a vector field $\xi^\mu$ is
\be
\delta_\xi \phi=\mathcal{L}_\xi \phi,
\ee
with the right hand side being the Lie derivative of $\phi$ with respect to $\xi$ \cite{Wald:1984rg}.  To make contact with the notation of the previous section we can define a vector field
\be\label{Xxidef}
X_\xi\equiv \int d^d x \mathcal{L}_{\xi} \phi^a(x)\frac{\delta}{\delta\phi^a}
\ee
on configuration space, in terms of which we have
\be
\delta_\xi \phi^a(x)=\mathcal{L}_{X_\xi}\phi^a(x)=X_\xi \cdot \delta \phi^a(x).
\ee
Here ``$\cdot$'' again denotes the insertion of a vector into the first (and in this case only) argument of a differential form.  More generally the infinitesimal diffeomorphism transformation of any configuration-space tensor $\sigma$, such as the one-forms $\Theta$ and $C$ or the two-form $\omega$, is given by
\be\label{xisigmadef}
\delta_\xi \sigma\equiv \mathcal{L}_{X_\xi}\sigma.
\ee
In particular from \eqref{cartan} we have
\be
\delta_\xi \delta \phi^a(x)=\delta(X_\xi\cdot \delta \phi^a(x))=\delta(\mathcal{L}_\xi\phi^a(x)),
\ee
so ``the diffeomorphism of a variation is the variation of a diffeomorphism'', as is the case for the standard interpretation of the symbol $\delta\phi^a(x)$ as an infinitesimal function.

We now introduce a key definition: a configuration-space tensor $\sigma$ which is also a spacetime tensor locally constructed out of the dynamical and background fields is \textit{covariant} under the infinitesimal diffeomorphism generated by a vector field $\xi^\mu$ if
\be\label{Linv}
\delta_\xi \sigma = \mathcal{L}_\xi \sigma,
\ee
where we emphasize that $\mathcal{L}_\xi$ is the \textit{spacetime} Lie derivative. This is to be distinguished from the configuration-space Lie derivative $\mathcal{L}_{X_\xi}$ appearing in \eqref{xisigmadef}: the latter implements the diffeomorphism on dynamical fields only, while the former implements it on both dynamical and background fields.  This distinction is important because symmetries are only allowed to act on dynamical fields, so for $\sigma$ to transform correctly under a diffeomorphism symmetry it must be covariant.

The simplest way for a configuration space and spacetime tensor $\sigma$ locally constructed out of dynamical and background fields to be covariant under some $\xi$ is for all background fields involved in its construction to be invariant under $\xi$, in the sense that $\mathcal{L}_\xi \chi^i=0$ where $i$ runs over background fields $\chi^i$.   For example the Lagrangian form \eqref{scalarL} is covariant under any diffeomorphisms which are isometries of the background metric $g$, but it is not covariant under general diffeomorphisms.  More generally some non-invariant background fields are allowed as long as the combinations in which they appear in $\sigma$ are invariant.  An extreme case is for $\sigma$ to not depend on any nontrivial background fields at all, as happens for the Einstein-Hilbert Lagrangian in general relativity, in which case it will be covariant under \textit{arbitrary} diffeomorphisms.\footnote{A trivial background field is one which is invariant under arbitrary diffeomorphisms.  One example is a coupling constant, and another is the $\epsilon$ symbol.}  In fact it was shown in \cite{Iyer:1994ys} that this is the only way for a Lagrangian form to be covariant under arbitrary diffeomorphisms: it must be built only out of a dynamical metric $g_{\mu\nu}$, its associated Riemann tensor $R^\alpha_{\phantom{\alpha}\beta\gamma\delta}$, tensorial dynamical matter fields, and covariant derivatives of the latter two.\footnote{One small exception is that the Lagrangian in this form may be entirely independent of the metric, as happens e.g. in Chern-Simons theory, in which case we do not need the metric to be dynamical.  Also \cite{Iyer:1994ys} did not consider spinor fields or connections on nontrivial bundles, but their argument was generalized to include them in \cite{Prabhu:2015vua}.  In this paper we are not explicitly addressing such fields, but we expect they can be included in our formalism with little modification.}  Moreover it was also shown that for such Lagrangians the symplectic potential $\Theta$ can always be taken to be covariant under arbitrary diffeomorphisms, essentially because the derivation of \eqref{Lvar} can always be done using ``integration by parts'' manipulations on covariant derivatives.  Indeed even if there are nontrivial background fields, we can still choose $\Theta$ to be covariant under the subgroup of diffeomorphisms which preserve all background fields.   This is because we could always choose to consider a different theory where all background fields become dynamical, in which case the Lagrangian form would become covariant under arbitrary diffeomorphisms, and thus by the argument of \cite{Iyer:1994ys} so would $\Theta$.  Therefore $\Theta$ must still be covariant in the original theory under diffeomorphisms which preserve all background fields.

Covariance of the Lagrangian form $L$ under the diffeomorphisms generated by a vector field $\xi^\mu$ is not sufficient for those diffeomorphisms to be symmetries. For a continuous transformation of dynamical fields to be a symmetry, this transformation must respect the boundary conditions and the action must be invariant under that transformation up to possible boundary terms at $\Sigma_{\pm}$ (see section \ref{noethersec} below for more on why this is the correct requirement).  These requirements are nontrivial, for example many diffeomorphisms do not even preserve the location of $\Gamma$.    We can write the variation of the action \eqref{Sdef} by an infinitesimal diffeomorphism under which $L$ is covariant as
\begin{align}\nonumber
\delta_\xi S=&\int_M \delta_\xi L+\int_{\partial M}\delta_\xi \ell\\
=&\int_{\partial M} \left(\xi\cdot L+\delta_\xi \ell\right),\label{Sdiff}
\end{align}
where we have used \eqref{Linv} and \eqref{cartan}.  To avoid contributions at the spatial boundary $\Gamma$, we first require that at $\Gamma$ the normal component of $\xi^\mu$ vanishes.  This ensures that $\xi^\mu$ does not move $\Gamma$, and also ensures that the first term in \eqref{Sdiff} vanishes.  We then also require that $\ell$ be covariant with respect to $\xi$: in this case the second term also does not give a contribution at $\Gamma$, since we then have $\delta_\xi \ell|_\Gamma=\mathcal{L}_\xi \ell|_{\Gamma}=d(\xi\cdot \ell)|_\Gamma$, which integrates to an allowed contribution at $\partial \Sigma_{\pm}$.  In general this covariance of $\ell$ imposes more requirements on $\xi$ than just a vanishing normal component at $\Gamma$. We thus will need to restrict consideration to diffeomorphisms obeying these additional requirements (and also preserving the boundary conditions), since otherwise they will not be symmetries.

In considering what kinds of terms may appear in $\ell$ it is useful to adopt the covariant hypersurface formalism, which is a way of discussing the extrinsic properties of a hypersurface without making any choice of coordinates \cite{Wald:1984rg,Carroll:2004st}.  To discuss $\partial M$ in this formalism, we introduce a background scalar field $f$ on $M$ such that
\bi
\item[(1)] There is a neighborhood of $\partial M$ in which $f\leq 0$, and in which $f=0$ only on $\partial M$.
\item[(2)] $\partial_\mu f$ is either spacelike or timelike at each point in $\partial M$, except perhaps at finitely many ``corners'' where it is not well-defined and across which its signature can switch.
\ei
Different choices of $f$ away from $\partial M$ give different foliations of the spacetime near the boundary.  We can then define a normal one-form field
\be\label{normaldef}
n_\mu\equiv\frac{\partial_\mu f}{\sqrt{\pm \partial_\alpha f\partial_\beta f g^{\alpha \beta}}}
\ee
in the vicinity of $\partial M$, with the $\pm$ being determined by whether $\partial_\mu f$ is spacelike or timelike on the nearby part of $\partial M$.\footnote{\label{cornernote}This notion is ambiguous in the vicinity of a corner where the signature of $\partial_\mu f$ changes sign, in what follows the values of any quantities at such corners are always defined by approaching them from the spatial boundary $\Gamma$.  Also we note that this (standard) definition has the somewhat counter-intuitive property that if $n_\mu$ is timelike and $f$ is increasing towards the future, then $n^\mu$ is past-pointing.}     $n_\mu$ can then be used to define an induced metric
\be\label{gammadef}
\gamma_{\mu\nu}\equiv g_{\mu\nu}\mp n_\mu n_\nu
\ee
and an extrinsic curvature tensor
\be\label{Kdef}
K_{\mu\nu}=\frac{1}{2}\mathcal{L}_n \gamma_{\mu\nu}=\gamma_\mu^{\phantom{\mu}\alpha}\nabla_\alpha n_\nu,
\ee
where we emphasize that these quantities live in a neighborhood of $\partial M$.  Away from $\partial M$ in this neighborhood they obviously depend on the choice of $f$, but right on $\partial M$ they do not.\footnote{To see this, note that if $f$ and $f'$ both vanish on $\partial M$, with both of their gradients having the same signature, then we must have $f'=hf$, with $h$ some scalar function which is nonvanishing on $\partial M$.  But then on $\partial M$ we have $\partial_\mu f'=h\partial_\mu f$, so they define the same $n_\mu$ there.  $\gamma_{\mu\nu}$ will then also be the same, and so will $K_{\mu\nu}$ since the second equality in \eqref{Kdef} makes it clear that to define $K_{\mu\nu}$ we only need to differentiate $n_\mu$ ``along'' $\partial M$.}  This neighborhood will be foliated by slices of constant $f$, and within it $\gamma_{\mu}^{\phantom{\mu}\nu}$ can be used to project tensor indices down to ones which are tangent to those slices.  It can also be used to define a hypersurface-covariant derivative, which, acting on any tensor $T$ that obeys the requirement that contraction of any index with $n_\mu$ or $n^\mu$ vanishes, is defined by
\be\label{Ddef}
D_\mu T^{\alpha_1\ldots \alpha_m}_{\phantom{\alpha_1\ldots \alpha_m}\beta_1\ldots \beta_n}\equiv \gamma_\mu^{\phantom{\mu}\nu}\gamma_{\sigma_1}^{\phantom{\sigma_1}\alpha_1}\ldots \gamma_{\sigma_m}^{\phantom{\sigma_m}\alpha_m} \gamma_{\beta_1}^{\phantom{\beta_1}\\\rho_1}\ldots \gamma_{\beta_n}^{\phantom{\beta_n}\rho_n} \nabla_\nu T^{\sigma_1\ldots \sigma_m}_{\phantom{\sigma_1\ldots \sigma_m}\rho_1\ldots \rho_n}.
\ee
This is the unique derivative such that $D_\mu \gamma_{\alpha\beta}=0$.  $\gamma_{\mu\nu}$, $n_\mu$, and $K_{\mu\nu}$ (and also their tangential and normal derivatives) are natural quantities to use in constructing $\ell$, together with tangential and normal derivatives of the dynamical fields.

By construction, $\ell$ will transform as a $(d-1)$-form under diffeomorphisms which act on both dynamical and background fields, with $f$ included among the latter.  For it to be covariant we need it to still transform as a $(d-1)$-form when only the dynamical fields transform.  We've already seen that the covariance of $L$ under the infinitesimal diffeomorphisms generated by $\xi^\mu$ requires all `bulk'' background fields to appear in $L$ only in combinations which are invariant under those diffeomorphisms.  Similarly the covariance of $\ell$ also requires some kind of invariance of $f$.  We only need $\ell$ to be covariant at the spatial boundary $\Gamma$, so the strongest condition we could reasonably require is that
\be
\xi^\nu \partial_\nu f=0
\ee
everywhere in some neighborhood of $\Gamma$, in which case we will will say that $\xi^\mu$ is \textit{foliation-preserving}.  $\ell$ will always be covariant with respect to foliation-preserving diffeomorphisms (provided that any other background fields are also invariant).  More generally however we can also consider diffeomorphisms where we only require
\be\label{orderk}
n^{\mu_1}\ldots n^{\mu_n}\nabla_{\mu_1}\ldots \nabla_{\mu_n}\left(\xi^\nu n_\nu\right)|_{\Gamma}=0
\ee
for all $n=0,1,\ldots k$, in which case we say that $\xi^\mu$ is \textit{foliation-preserving at order k}.  Any $\ell$ which is constructed out of at most $k$ derivatives of $f$ will also be covariant under such diffeomorphisms,\footnote{Indeed note that if $f$ were dynamical, we would have $\delta_\xi\partial_{\mu_1}\ldots \partial_{\mu_n}f=\partial_{\mu_1}\ldots \partial_{\mu_n}(\xi^\nu \partial_\nu f)$.  In fact $\partial_{\mu_1}\ldots \partial_{\mu_n} f$ is only a background field, and thus should not transform, but we can still preserve covariance provided that $\partial_{\mu_1}\ldots \partial_{\mu_n}(\xi^\nu \partial_\nu f)=0$ for all $n\leq k$, which is equivalent to \eqref{orderk} holding for all $n\leq k$.} and in fact since $f$ appears only inside of $n_\mu$, which is foliation-independent, such an $\ell$ will actually also be covariant under foliation-preserving diffeomorphisms of order $k-1$.\footnote{This is because if more than one derivative acts on $f$ there will always be at least one which is taken parallel to the foliation. }

Finally we consider the covariance of the quantity $C$ appearing in \eqref{Ceq}.  We will assume that given $\ell$ and $\Theta$ the demonstration of equation \eqref{Ceq} involves ``covariant integration by parts'' manipulations on the boundary, together with imposing the boundary conditions (see sections \ref{scalarsec}, \ref{grsec} for examples of this).  The $C$ which appears will then always be a locally constructed out of the dynamical and background fields and their derivatives, and it will transform as a $(d-2)$-form under diffeomorphisms which act on both the dynamical and background fields.  Moreover like $\ell$ it will be covariant under foliation-preserving diffeomorphisms which preserve any other background fields.  Furthermore if $\ell$ involves at most $k$ derivatives of $f$ then $C$ will as well, so $C$ will more generally at least be covariant under foliation-preserving diffeomorphisms of order $k-1$.  We will need to use this covariance of $C$ in the following subsection.

\subsection{Diffeomorphism charges}\label{chargesec}
We now turn to the problem of constructing the Hamiltonian $H_\xi$ that generates the evolution in phase space corresponding to the diffeomorphisms generated by any vector field $\xi^\mu$ which respects the boundary conditions and under which $L$, $\ell$, and $C$ are covariant.  Our strategy will be to first find a function $H_\xi$ on pre-phase space obeying
\be\label{inverseH}
\delta H_\xi=-X_\xi\cdot \wt{\Omega},
\ee
with $X_\xi$ given by \eqref{Xxidef}.  For any zero mode $\wt{X}$ of $\wt{\Omega}$ we have
\be
\wt{X}\cdot \delta H_\xi=\wt{\Omega}(\wt{X},X_\xi)=0,
\ee
so $H_\xi$ will also be a well-defined function on the phase space $\Pc$. Moreover since $\wt{\Omega}$ defines the non-degenerate symplectic form $\Omega$ on $\Pc$ via \eqref{Omegadefgen}, we may use its inverse there to rewrite \eqref{inverseH} as
\be
X_\xi(f)=\Omega^{-1}(\delta f,\delta H_{\xi}),
\ee
where $X_\xi$ is now defined modulo addition by a zero mode of $\wt{\Omega}$ and $f$ is a function on $\Pc$.  This is nothing but Hamilton's equation \eqref{XHeq}, so finding an $H_\xi$ on $\wt{\Pc}$ obeying \eqref{inverseH} is sufficient to construct the Hamiltonian on phase space.

We now compute the right hand side of \eqref{inverseH}, aiming to show that indeed it is equal to $\delta$ of something.  It is useful \cite{Iyer:1994ys} to first introduce the Noether current
\be\label{noetherdef}
J_\xi\equiv X_{\xi}\cdot \Theta-\xi\cdot L.
\ee
This is a scalar function on $\wt{\Pc}$, and a $(d-1)$-form on spacetime.  Note that we are using ``$\cdot$'' for the insertion of both pre-phase space and spacetime vectors.  If $L$ is covariant under $\xi$ then $J_\xi$  is closed as a spacetime form:
\begin{align}\nonumber
dJ_\xi&=d(X_{\xi}\cdot\Theta)-d(\xi\cdot L)\\\nonumber
&=X_\xi\cdot\left(\delta L-E_a\delta\phi^a\right)-\mathcal{L}_\xi L\\\nonumber
&=\delta_\xi L-\mathcal{L}_\xi L-E_a \mathcal{L}_\xi \phi^a\\
&=0.
\end{align}
In this derivation we have used \eqref{Lvar}, \eqref{cartan}, \eqref{EOM}, \eqref{Linv}, and also that $d(X_\xi\cdot \Theta)=X_\xi\cdot d\Theta$.  We then have the following calculation:
\begin{align}\nonumber
-X_\xi\cdot \omega&=-X_\xi\cdot \delta(\Theta-dC)\\\nonumber
&=\delta\left(X_\xi\cdot(\Theta-dC)\right)-\mathcal{L}_{X_\xi}(\Theta-dC)\\\nonumber
&=\delta J_\xi+\xi\cdot \delta L-\mathcal{L}_\xi \Theta+d\left(\delta_\xi C-\delta(X_\xi\cdot C)\right)\\\nonumber
&=\delta J_\xi+\xi\cdot(d\Theta+E_a\delta\phi^a)-\mathcal{L}_\xi\Theta+d\left(\delta_\xi C-\delta(X_\xi\cdot C)\right)\\
&=\delta J_\xi+d\left(\delta_\xi C-\delta(X_\xi\cdot C)-\xi\cdot \Theta\right).\label{Xomega}
\end{align}
Here we have made liberal use of \eqref{cartan} for both pre-phase space and spacetime differential forms, as well as \eqref{noetherdef}, \eqref{Lvar}, \eqref{xisigmadef}, \eqref{Linv} (applied to $\Theta$), and \eqref{EOM}.  We have not yet applied \eqref{Linv} to $C$, since we are only assuming that $C$ is covariant at $\Gamma$.  We may do so after integrating over a Cauchy slice $\Sigma$, to obtain
\begin{align}\nonumber
-X_\xi\cdot \wt{\Omega}&=\int_\Sigma \delta J_\xi+\int_{\partial \Sigma}\left(\mathcal{L}_\xi C-\delta(X_\xi\cdot C)-\xi\cdot \Theta\right)\\\nonumber
&=\int_\Sigma \delta J_\xi+\int_{\partial \Sigma}\left(\xi\cdot (dC-\Theta)-\delta(X_\xi\cdot C)\right)\\
&=\delta\left(\int_\Sigma J_\xi+\int_{\partial \Sigma}\left(\xi\cdot \ell-X_\xi\cdot C\right)\right).
\end{align}
Here we have again used \eqref{cartan}, as well as \eqref{Linv} (applied to $C$) and \eqref{Ceq}, and also discarded the integral of a total derivative over the closed manifold $\partial \Sigma$.  Comparing to \eqref{inverseH} we see that we have succeeded in obtaining an exterior derivative on pre-phase space, with $H_\xi$ given by
\be\label{Hresult1}
H_\xi\equiv\int_\Sigma J_\xi+\int_{\partial \Sigma}\left(\xi\cdot \ell-X_\xi\cdot C\right)+\mathrm{constant},
\ee
where the arbitrary additive constant is independent of the dynamical fields and reflects the standard additive ambiguity of the energy in any Hamiltonian system.  Note in particular that no ``integrability condition'', such as those in equation (80) of \cite{Iyer:1994ys}  or equation (16) of \cite{Wald:1999wa}, needed to be introduced during this derivation: equation \eqref{Ceq}, which we obtained by demanding stationarity of the action up to future/past terms, was sufficient to algorithmically construct $H_\xi$.\footnote{This is not to say that equation (16) of \cite{Wald:1999wa}, or a version of equation (80) from \cite{Iyer:1994ys} which accounts for $C\neq 0$, does not hold: they do hold, but they are consequences of our assumption that the variational problem is well-posed rather than additional assumptions.}  Note also that $H_\xi$ is independent of choice of Cauchy surface $\Sigma$: if we consider two slices $\Sigma'$ and $\Sigma$, whose boundaries obey $\partial \Sigma'-\partial \Sigma=\partial \Xi$, with $\Xi\subset \Gamma$, the difference of $H_\xi$ evaluated on these slices is given by
\begin{align}\nonumber
\int_{\Xi}\left(J_\xi+d(\xi\cdot \ell-X_\xi\cdot C)\right)&=\int_\Xi\left(X_\xi\cdot (\Theta-dC)-\xi\cdot L+d(\xi\cdot \ell)\right)\\\nonumber
&=\int_\Xi\left(-X_\xi\cdot \delta \ell+d(\xi\cdot\ell)-\xi\cdot L\right)\\\nonumber
&=\int_\Xi\left(-\delta_\xi \ell+\mathcal{L}_\xi \ell-\xi\cdot L\right)\\
&=0.
\end{align}
Here we used \eqref{noetherdef}, \eqref{Ceq}, \eqref{xisigmadef} applied to $\ell$, \eqref{cartan}, \eqref{Linv} applied to $\ell$, and that $\xi$ has no normal component to $\Xi$.

Our derivation of \eqref{Hresult1} only required the various quantities to be covariant with respect to the particular diffeomorphism $\xi^\mu$ being considered.  So for example we could use \eqref{Hresult1} to write down the various Poincare generators of any relativistic Lagrangian field theory in Minkowski space.  In the special case where $L$ is covariant under \textit{arbitrary} continuous diffeomorphisms, as happens for example in general relativity, an additional simplification of \eqref{Hresult1} is possible.  Indeed in this situation it was shown in \cite{wald1990identically,Iyer:1994ys} that not only do we have $dJ_\xi=0$, actually there will be a local covariant $(d-2)$-form $Q_\xi$ constructed out of the dynamical fields and their derivatives, called the \textit{Noether charge}, such that\footnote{This name is rather misleading: $Q_\xi$ is not conserved and does not generate any symmetry.  ``Noether potential'' would have been better, it is $H_\xi$ which is really the Noether charge.}
\be\label{JdQ}
J_\xi=dQ_\xi.
\ee
We may then make one final application of Stokes theorem in \eqref{Hresult1} to obtain the following expression, true only in generally-covariant theories:
\be\label{Hresult2}
H_\xi=\int_{\partial \Sigma}\left(Q_\xi+\xi\cdot \ell-X_\xi\cdot C\right)+\mathrm{constant}.
\ee
Thus in such theories the Hamiltonian for any continuous diffeomorphism is a pure boundary term: this is analogous to the fact that in electromagnetism that the total electric charge is the electric flux through spatial infinity.

Equations \eqref{Hresult1} and \eqref{Hresult2} are perhaps the main technical results of this paper; as far as we know they have not appeared in the literature before.  One can obtain equation (82) from reference \cite{Iyer:1994ys} by replacing $\ell \to -B$ and setting $C=0$: the terms involving $C$ are not present there because $C$ was not included in their definition of the pre-symplectic current, while we included it in \eqref{omegadef} to ensure that $\omega|_{\Gamma}=0$.\footnote{In \cite{Iyer:1994ys} the possibility of such a modification of $\omega$ was considered in the discussion around equations (46-48), but dismissed basically on grounds that $C$ would be hard to extend covariantly into $M$.  A covariant extension is not necessary however, and indeed the $C$ terms we construct in sections \ref{scalarsec} and \ref{grsec} do not have one.}

The boundary terms in \eqref{Hresult1} can be given a nice interpretation as follows.  As mentioned in footnote \ref{redef}, if we are not interested in preserving the covariance of $L$ and $\Theta$ then we can remove the boundary term $\ell$ from the action and the total derivative $dC$ from equation \eqref{Ceq} via the redefinitions
\begin{align}\nonumber
L'&\equiv L+d\ell\\
\Theta'&\equiv\Theta+\delta \ell-dC.
\end{align}
In terms of these the action and presymplectic current are simply
\begin{align}\nonumber
S&=\int_M L'\\
\omega&=\delta \Theta'
\end{align}
We can also define a new Noether current
\begin{align}\nonumber
J'_\xi&\equiv X_\xi\cdot\Theta'-\xi\cdot L'+(\mathcal{L}_\xi-\delta_\xi)\ell\\
&=J_\xi+d(\xi\cdot \ell-X_\xi\cdot C),
\end{align}
where the extra terms involving $\ell$ in the definition are necessary to ensure that $dJ_\xi'=0$.\footnote{These terms also follow from the general Noether theorem we present in section \ref{noethersec} below, as we will explain there.} We thus may rewrite \eqref{Hresult1} as
\be\label{Hpdef}
H_\xi=\int_\Sigma J'_{\xi},
\ee
so we see that it is really $J'_\xi$ which should be thought of as the local generator of $\xi$ diffeomorphisms.  Moreover if we choose $f$ away from $\partial M$ such that $\xi$ is foliation-preserving near $\Sigma$, then the $\ell$ terms in the definition of $J'_\xi$ do not contribute to $H_\xi$.  We then have
\be
H_\xi=\int_\Sigma\left(X_\xi\cdot \Theta'-\xi\cdot L'\right),
\ee
which is a version of the standard formula $H=p\dot{q}-L$.

There is an aspect of this construction which may seem mysterious: it has not required any discussion of whether or not the equations of motion have a well-posed initial value formulation in the usual sense of fixing some data about the dynamical fields and their derivatives on a Cauchy slice $\Sigma$ and asking for a unique time evolution off of the slice (see e.g. \cite{Wald:1984rg}).  In the usual non-covariant way of thinking about the Hamiltonian formalism, one defines phase space in terms of the values of the dynamical fields and some number of their derivatives on a Cauchy slice.  A well-posed initial value formulation is then necessary in order for the Hamiltonian and the symplectic form to be sufficiently smooth objects on this phase space that the relevant theorems on vector flows ensure a good dynamics.  The reason we have not encountered the initial value problem in our formalism is that we have simply \textit{defined} the pre-phase space $\wt{\Pc}$ to be the set of field configurations which obey the equations of motion throughout spacetime, so any initial data we find on any Cauchy slice must be of the type which allows such a solution.  In theories which do have a satisfactory initial value formulation, points in our phase space $\Pc=\wt{\Pc}/\wt{\mathcal{G}}$ will be in one-to-one correspondence with some natural set of initial data on each Cauchy slice.  More generally, our $\Pc$ will only be some subset of the ``naive'' phase space one might construct on a Cauchy slice by studying the constraints and the number of derivatives.  In fact for poor choices of theory there may be no solutions at all!  Identifying the set of initial data which leads to valid solutions is a problem about which can say little in general, as we expect it to depend on the details of both the Lagrangian and the boundary conditions.  We do expect however that on our phase space $\mathcal{P}$ the Hamiltonian \eqref{Hresult1} and symplectic form \eqref{Omegadef} are sufficiently smooth to generate the expected dynamics, and moreover we expect that if $L$, $\ell$, and the boundary conditions depend on only finitely many derivatives of the dynamical fields, then the values of the dynamical fields and some finite number of their derivatives on a Cauchy slice are sufficient to determine a unique solution up to gauge transformations, provided that any solution exists with that initial data.

\section{Examples}
We now illustrate this formalism in a series of examples, starting simple to get some practice with our differential form technology.
\subsection{Particle mechanics}
We first consider the mechanics of $n$ particles with positions $q^a$ and Lagrangian form
\be
L=\mathcal{L}(q^a,\dot{q}^a)dt.
\ee
The variation of this Lagrangian form is
\be
\delta L=\left(\frac{\partial \mathcal{L}}{\partial q^a}-\frac{d}{dt}\frac{\partial \mathcal{L}}{\partial \dot{q}^a}\right)dt\delta q^a+d\Theta,
\ee
with
\be
\Theta=\frac{\partial \mathcal{L}}{\partial \dot{q}^a}\delta q^a.
\ee
The Noether current for the time translation $\xi=\frac{d}{dt}$ is
\be
J_\xi=\frac{\partial \mathcal{L}}{\partial \dot{q}^a}\dot{q}^a-\mathcal{L},
\ee
so defining $p_a\equiv \frac{\partial \mathcal{L}}{\partial \dot{q}^a}$ we arrive at the usual formula 
\be
H=p_a\dot{q}^a-\mathcal{L}
\ee
for the Hamiltonian in particle mechanics.  Similarly the symplectic form is
\be
\Omega=\delta\omega = \delta p_a\wedge \delta q^a.
\ee

\subsection{Two-derivative scalar field}
We next consider the scalar field theory \eqref{scalarL}, with Lagrangian form
\be
L=-\left(\frac{1}{2}\nabla_\mu \phi \nabla^\mu \phi+V(\phi)\right)\epsilon.
\ee
The variation of $L$ is
\be
\delta L = (\nabla_\mu \nabla^\mu\phi-V'(\phi))\epsilon\, \delta \phi+d\Theta,
\ee
with
\be
\Theta=\theta \cdot \epsilon,
\ee
where we define
\be
\theta^\mu\equiv -\nabla^\mu \phi\delta \phi,
\ee
and we have used the convenient identity
\be\label{extdiv}
d(V\cdot \epsilon)=\left(\nabla_\mu V^\mu\right)\epsilon,
\ee
which is true for any vector field $V$.  The restriction of $\Theta$ to $\partial M$ is given by
\be
\Theta|_{\partial M}=n_\mu \theta^\mu \epsilon_{\partial M},
\ee
where $\epsilon_{\partial M}$ is the volume form on $\partial M$, $n_\mu$ is the normal form \eqref{normaldef}, and we have used \eqref{epsilonrelation}.  Therefore our boundary requirement \eqref{Ceq} will be satisfied with $\ell$, $C=0$ provided that we adopt either Dirichlet ($\delta\phi|_{\Gamma}=0$) or Neumann ($n^\mu \nabla_\mu\phi|_{\Gamma}=0$) boundary conditions.  To write the pre-symplectic current we need to address a notational subtlety we have so far avoided: with two kinds of differential forms, there are also two kinds of wedge products.  We will from here on adopt a convention where we automatically view the configuration-space differentials $\delta\phi^a$ as anti-commuting objects.  The product of two of them will therefore implicitly be a wedge product, but we will only write ``$\wedge$'' for the spacetime wedge product.  With this convention, the pre-symplectic current is given by
\be
\omega=\delta \Theta=\hat{\omega}\cdot \epsilon,
\ee
with
\be
\hat{\omega}^\mu=-\nabla^\mu \delta \phi \,\delta \phi,
\ee
and the pre-symplectic form is
\be
\wt{\Omega}=\int_\Sigma \omega=\int_\Sigma \left(\hat{n}_\mu \hat{\omega}^\mu\right)\epsilon_\Sigma=-\int_\Sigma \left(\hat{n}^\mu\nabla_\mu \delta \phi \,\delta \phi\right)\epsilon_\Sigma.
\ee
Here $\hat{n}^\mu$ is the normal vector to $\Sigma$, which we note is past-pointing in our conventions (see the discussion around \eqref{epsilonrelation}).  This pre-symplectic form is already non-degenerate, so no quotient is necessary and we have $\Pc=\wt{\Pc}$ and $\Omega=\wt{\Omega}$.  Indeed comparing to \eqref{pqomega}, we see that we have recovered using covariant methods the standard result that in this theory $\dot{\phi}$ is the canonical momentum conjugate to $\phi$.  Finally the Noether current is
\be
J_\xi=j_\xi\cdot \epsilon,
\ee
with
\be
j_\xi^\mu=-\xi_\nu\left(\nabla^\mu\phi\nabla^\nu \phi-g^{\mu\nu}\left(\frac{1}{2}\nabla_\alpha\phi\nabla^\alpha \phi+V(\phi)\right)\right),
\ee
where the quantity in brackets is the energy-momentum tensor $T^{\mu\nu}$.  $J_\xi$ is closed on $\wt{\Pc}$ if and only if $\xi^\mu$ is a Killing vector of the background metric.

\subsection{Maxwell theory}\label{maxwellsec}
We now give an example where the quotient from pre-phase space to phase space is nontrivial.  This will just be Maxwell electrodynamics, with Lagrangian form
\be
L=-\frac{1}{2} F\wedge \star F.
\ee
Its variation is
\be
\delta L=-\delta A\wedge d\star F-d\left(\delta A\wedge \star F\right),
\ee
so apparently we have
\be
\Theta=-\delta A\wedge \star F.
\ee
If we impose Dirichlet boundary conditions, meaning we fix the pullback of $A$ to the spatial boundary $\Gamma$, then the stationarity requirement \eqref{Ceq} is satisfied with no need for an $\ell$ or $C$.    We then have the symplectic potential
\be
\omega\equiv \delta \Theta=\delta A \wedge \star \delta F,
\ee
and pre-symplectic form
\be
\wt{\Omega}=\int_\Sigma\left(\delta A \wedge \star \delta F\right),
\ee
which illustrate the usual statement that $A$ and $-\star F$ are canonical conjugates.  Zero modes of $\wt{\Omega}$ are associated with gauge transformations, which are flows in configuration space generated by vectors of the form
\be
X_\lambda\equiv \int d^d x \partial_\mu \lambda \frac{\delta}{\delta A_\mu}.
\ee
Indeed note that
\begin{align}\nonumber
X_\lambda\cdot \wt{\Omega}&=\int_\Sigma\left(d\lambda \wedge \star \delta F\right)\\\nonumber
&=\int_\Sigma d\left(\lambda \star \delta F\right)\\
&=\int_{\partial \Sigma}\lambda \star \delta F.\label{Xomegamaxwell}
\end{align}
Our Dirichlet boundary conditions require the restriction of $d\lambda$ to $\Gamma$ vanishes, so $\lambda$ must be constant on $\Gamma$.  Since the boundary conditions allow for $\int_{\partial \Sigma} \star F$ to vary, $X_\lambda$ will apparently be a zero mode of $\wt{\Omega}$ if and only if $\lambda|_{\Gamma}=0$.  Therefore in constructing the physical phase space we should quotient only by the set of gauge transformations which vanish at the spatial boundary.  The ones which approach a nonzero constant there act nontrivially on phase space, and in fact by an analogue of the discussion below \eqref{inverseH} we can interpret \eqref{Xomegamaxwell} as telling us that the generator of these gauge transformations on phase space is\footnote{We have switched the sign here compared to \eqref{inverseH} to respect the standard convention that in quantum mechanics a time translation is $e^{-iHt}$ while an internal symmetry rotation is $e^{i\lambda Q}$.}
\be
Q_\lambda\equiv \lambda\int_{\partial \Sigma}\star F,
\ee
as expected from Gauss's law.

\subsection{Higher-derivative scalar}\label{scalarsec}
We now give a simple example of a theory with nonzero $C$.  This is a non-interacting scalar field theory, with Lagrangian form
\be
L=-\frac{1}{2}\left(\nabla_\mu \phi \nabla^\mu \phi+\nabla_\mu\nabla_\nu\phi \nabla^\mu\nabla^\nu\phi\right)\epsilon.
\ee
We first note that
\be
\delta L=\left(\nabla_\mu\nabla^\mu \phi-\nabla_\mu\nabla_\nu \nabla^\nu\nabla^\mu\phi\right)\,\epsilon\,\delta \phi+d\Theta,
\ee
with
\be
\Theta=\theta\cdot \epsilon,
\ee
with $\theta$ being the vector
\be
\theta^\mu\equiv \left(\nabla_\nu\nabla^\nu\nabla^\mu \phi-\nabla^\mu \phi\right)\delta \phi-\nabla^\mu\nabla^\nu\phi \nabla_\nu \delta \phi.
\ee

To identify $C$ we are interested in the pullback of $\Theta$ to the $\partial M$, which from \eqref{epsilonrelation} is given by
\be
\Theta|_{\partial M}=\theta^\mu n_\mu\epsilon_{\partial M}.
\ee
We will show that $\Theta |_{\partial M}$ is the sum of a term which vanishes with appropriate boundary conditions and a term which is a boundary total derivative.  Indeed by using \eqref{gammadef} to decompose the $\nabla_\nu \delta\phi$ in the third term of $\theta^\mu$ into normal and tangential parts, we find
\begin{align}\nonumber
\theta^\mu n_\mu=&\left(n^\mu\left(\nabla^\nu\nabla_\nu\nabla_\mu\phi-\nabla_\mu\phi\right)+D_\alpha\left(\gamma^{\alpha\beta}n^\mu \nabla_\mu\nabla_\beta\phi\right)\right)\delta\phi\\\nonumber
&\mp \left(n^\mu n^\alpha  \nabla_\mu\nabla_\alpha\phi\right) n^\beta\nabla_{\beta}\delta \phi\\
&-D_\alpha\left(\gamma^{\alpha\beta}n^\mu \nabla_\mu \nabla_\beta\phi \,\delta \phi\right).
\end{align}
Here $D_\alpha$ is the hypersurface-covariant derivative \eqref{Ddef}.  Therefore if we adopt ``generalized Neumann'' boundary conditions
\begin{align}\nonumber
n^\mu\left(\nabla^\nu\nabla_\nu\nabla_\mu\phi-\nabla_\mu\phi\right)|_{\Gamma}+D_\alpha\left(\gamma^{\alpha\beta}n^\mu \nabla_\mu\nabla_\beta\phi\right)|_{\Gamma}&=0\\
n^\mu n^\alpha  \nabla_\mu\nabla_\alpha\phi|_{\Gamma}&=0,
\end{align}
then \eqref{Ceq} holds provided we define
\be
C\equiv c\cdot \epsilon_{\partial M},
\ee
with
\be
c^\mu\equiv -\gamma^{\mu\alpha}n^\beta \nabla_\alpha\nabla_\beta\phi \,\delta\phi.
\ee
This $C$ term is not covariant in the interior of $M$, but by the discussion above equation \eqref{orderk} its restriction to the boundary will be covariant under foliation-preserving diffeomorphisms of order zero.

We expect this example is indicative of the general situation for higher derivative Lagrangians: there will typically be a nonvanishing $C$ term, which is covariant on the boundary but cannot be covariantly extended into the interior of $M$.

\subsection{General relativity}\label{grsec}
We now discuss general relativity, which we take to have
\begin{align}\nonumber
L&=\frac{1}{16\pi G}\left(R-2\Lambda\right)\epsilon \\
\ell&=\frac{1}{8\pi G}K \,\epsilon_{\partial M}.\label{GRact}
\end{align}
Here $R$ is the Ricci scalar, and $K$ is the trace $g^{\alpha\beta}K_{\alpha\beta}$ of the extrinsic curvature \eqref{Kdef}.  The metric $g_{\mu\nu}$ is dynamical, and there are no nontrivial background fields.  The relevant variations (see e.g. \cite{Wald:1984rg}) are
\begin{align}\nonumber
\delta \epsilon&=\left(\frac{1}{2} g^{\mu\nu}\delta g_{\mu\nu}\right)\epsilon\\\nonumber
\delta \epsilon_{\partial M}&=\left(\frac{1}{2}\gamma^{\mu\nu}\delta g_{\mu\nu}\right)\epsilon_{\partial M}\\\nonumber
\delta\Gamma^\mu_{\alpha\beta}&=\frac{1}{2}g^{\mu\nu}\left(\nabla_\alpha \delta g_{\beta\nu}+\nabla_{\beta}\delta g_{\alpha \nu}-\nabla_\nu \delta g_{\alpha\beta}\right)\\\nonumber
\delta R&=-R^{\mu\nu}\delta g_{\mu\nu}+\nabla^\mu \nabla^\nu \delta g_{\mu\nu}-\nabla_\rho\nabla^\rho g^{\mu\nu}\delta g_{\mu\nu}\\\nonumber
\delta n_\mu&=\frac{1}{2}n^\alpha \left(\delta^{\beta}_{\phantom{\beta}\mu}-\gamma^{\beta}_{\phantom{\beta}\mu}\right)\delta g_{\alpha\beta}\\
\delta K&=-\frac{1}{2}K^{\mu\nu}\delta g_{\mu\nu}+\frac{1}{2}g^{\mu\nu}n^\lambda\nabla_\lambda \delta g_{\mu\nu}-\frac{1}{2}n^\alpha \nabla^\beta \delta g_{\alpha\beta}-\frac{1}{2}D_\mu\left(\gamma^{\mu\nu}n^\alpha \delta g_{\nu\alpha}\right),\label{pertgeom}
\end{align}
where $D_\mu$ is the hypersurface-covariant derivative \eqref{Ddef}, and we emphasize that in the last two variations we have treated the function $f$ identifying the location of $\partial M$ (see \eqref{normaldef}) as a background field. Using these variations we have
\be\label{GRvar}
\delta L=E^{\mu\nu}\delta g_{\mu\nu}+d\Theta,
\ee
with
\be
E^{\mu\nu}=\frac{1}{16\pi G}\left(-R^{\mu\nu}+\frac{1}{2}Rg^{\mu\nu}-\Lambda g^{\mu\nu}\right)\,\epsilon
\ee
and
\be
\Theta=\theta\cdot \epsilon,
\ee
where
\be\label{thetaeq}
\theta^\mu=\frac{1}{16\pi G}\left(g^{\mu\alpha}\nabla^\nu \delta g_{\alpha \nu}-g^{\alpha \beta}\nabla^\mu \delta g_{\alpha \beta}\right)
\ee
and we have used \eqref{extdiv}.  The equation of motion $E^{\mu\nu}=0$ is of course just the Einstein equation. Similarly we have the boundary variation
\be\label{ellvargr}
\delta \ell=\frac{1}{16\pi G}\left(\left(K\gamma^{\mu\nu}-K^{\mu\nu}\right)\delta g_{\mu\nu}+g^{\alpha\beta} n^\lambda\nabla_\lambda \delta g_{\alpha\beta}-n^\alpha\nabla^\beta\delta g_{\alpha\beta}-D_\mu\left(\gamma^{\mu\nu} n^\alpha \delta g_{\nu\alpha}\right)\right)\epsilon_{\partial M}.
\ee
Using \eqref{epsilonrelation}, the pullback of $\Theta$ to $\partial M$ is
\be
\Theta|_{\partial M}=n_\mu \theta^\mu\epsilon_{\partial M}.
\ee
Therefore from \eqref{thetaeq} and \eqref{ellvargr} we have
\be\label{GRbt}
\Theta|_{\partial M}+\delta \ell=-\frac{1}{16\pi G}\left(K^{\mu\nu}-K\gamma^{\mu\nu}\right)\epsilon_{\partial M}\delta g_{\mu\nu}+dC,
\ee
with
\be
C=c\cdot \epsilon_{\partial M}
\ee
and
\be\label{GRc}
c^\mu=-\frac{1}{16\pi G}\gamma^{\mu\nu}n^\alpha\delta g_{\nu\alpha}.
\ee
Thus \eqref{Ceq} will be satisfied provided that we choose boundary conditions such that
\be
\left(K^{\mu\nu}-K\gamma^{\mu\nu}\right)\delta g_{\mu\nu}|_\Gamma=0.
\ee
The boundary conditions we will adopt, analogous those we chose for Maxwell theory in section \ref{maxwellsec}, are to require that the pullback of $g_{\mu\nu}$ to $\Gamma$ is fixed.  We then must have
\be\label{GRbc}
\gamma_\mu^{\phantom{\mu}\alpha}\gamma_\nu^{\phantom{\nu}\beta}\delta g_{\alpha\beta}|_{\Gamma}=\gamma_\mu^{\phantom{\mu}\alpha}\gamma_\nu^{\phantom{\nu}\beta}\delta \gamma_{\alpha\beta}|_{\Gamma}=0.
\ee
The set of diffeomorphisms which respect \eqref{GRbc} are those for which $\xi^\mu n_\mu|_{\Gamma}=0$ and
\be
\gamma_\mu^{\phantom{\mu}\alpha}\gamma_\nu^{\phantom{\nu}\beta}\left(\nabla_\alpha\xi_\beta+\nabla_\beta \xi_\alpha\right)|_{\Gamma}=\left(D_\mu \xi_\nu+D_\mu \xi_\nu\right)|_{\Gamma}=0,
\ee
so in other words $\xi$ must approach a Killing vector of the spatial boundary metric.  In the language of section \ref{covlagsec} these diffeomorphisms are foliation-preserving at order zero, so since $\ell$ and $C$ are constructed out of $\gamma_{\mu\nu}$, $n_\mu$, and $K_{\mu\nu}$ they will be covariant.  With these boundary conditions $C$ is typically nonzero: $c^\mu$ involves the mixed normal-tangential components of $\delta g_{\nu\alpha}$, while \eqref{GRbc} only constrains the strictly tangential components.\footnote{\label{altbc}  We could also consider a stronger set of boundary conditions,  where \eqref{GRbc} is replaced by $\delta \gamma_{\mu\nu}|_\Gamma=0$.  We then would have to further restrict to diffeomorphisms obeying $n^\alpha \gamma_\mu^{\phantom{\mu}\beta}\left(\nabla_{\alpha}\xi_\beta+\nabla_\beta \xi_\alpha\right)|_\Gamma=0$.  Since $\gamma^{\mu\nu}\delta g_{\nu \lambda}=\gamma^{\mu\nu}\delta\gamma_{\nu \lambda}$, with these boundary conditions we would indeed have $C=0$.  Moreover the theory with these boundary conditions in fact is a partial gauge-fixing of the theory with the boundary conditions \eqref{GRbc}: we therefore construct the same physical phase space either way. This ability to get rid of $C$ with a partial gauge-fixing is special to general relativity, the theory of the previous subsection shows that it will not happen in general higher-derivative theories.}

We now consider the Noether current and charge.  From \eqref{noetherdef}, \eqref{GRact}, and \eqref{thetaeq} we have
\be
J_\xi=j_\xi\cdot \epsilon,
\ee
with
\be
j^\mu_\xi=\frac{1}{8\pi G}\left[\nabla_\nu\nabla^{[\nu}\xi^{\mu]}+\left(R^{\mu\nu}-\frac{1}{2}Rg^{\mu\nu}+\Lambda g^{\mu\nu}\right)\xi_\nu\right].
\ee
The results of \cite{wald1990identically,Iyer:1994ys} imply that on pre-phase space, where $E^{\mu\nu}=0$, we must have $J_\xi=dQ_\xi$ for some locally constructed $(d-2)$-form $Q_\xi$.  And indeed using the fact that for any two-form $S$ we have
\be
d\star S=s\cdot \epsilon\label{twoformswitch}
\ee
with
\be
s^\mu\equiv g^{\mu\alpha}\nabla^\beta S_{\alpha \beta},
\ee
we have
\be\label{GRQ}
Q_\xi=-\frac{1}{16\pi G}\star d \xi,
\ee
where we have viewed $\xi_\mu$ as a one-form.  More explicitly,
\be
(Q_\xi)_{\nu_1\ldots \nu_{d-2}}=-\frac{1}{16\pi G}\epsilon^{\alpha\beta}_{\phantom{\alpha\beta}\nu_1\ldots \nu_{d-2}}\nabla_\alpha\xi_\beta.
\ee

To compute $H_\xi$ we are interested in the pullback of $Q_\xi$ to $\partial \Sigma$, where $\Sigma$ is some Cauchy slice.  Constructing this is facilitated by observing that on $\Gamma$ we have
\be\label{epsilonpM}
\epsilon_{\partial M}=-\tau\wedge \epsilon_{\partial \Sigma},
\ee
where $\tau$ is the normal form of $\partial \Sigma$ viewed as the boundary of its past in $\Gamma$ (remember that this implies that $\tau^\mu$ is past-pointing).  The minus sign in \eqref{epsilonpM} follows from  the discussion of orientation below equation \eqref{stokes}.  Combining \eqref{epsilonrelation} and \eqref{epsilonpM} we have
\be
\epsilon=\tau\wedge n \wedge \epsilon_{\partial \Sigma},
\ee
so \eqref{GRQ} then gives
\be\label{GRQ2}
Q_\xi|_{\partial \Sigma}=-\frac{1}{16\pi G}\left(\tau^\alpha n^\beta-\tau^\beta n^\alpha\right)\nabla_\alpha \xi_\beta \,\epsilon_{\partial \Sigma}.
\ee
Similarly we have
\be
\xi\cdot \ell|_{\partial \Sigma}=-\frac{1}{8\pi G} \xi^\mu \tau_\mu K \epsilon_{\partial \Sigma}
\ee
and
\be\label{GRXC}
X_\xi\cdot C|_{\partial \Sigma}=\frac{1}{16\pi G}\left(\tau^\alpha n^\beta+\tau^\beta n^\alpha\right)\nabla_\alpha \xi_\beta \, \epsilon_{\partial \Sigma}.
\ee
Therefore from \eqref{Hresult2} we have
\begin{align}\nonumber
H_\xi&=-\frac{1}{8\pi G}\int_{\partial \Sigma} \left(\tau^\alpha n^\beta \nabla_{\alpha} \xi_\beta+\xi^\alpha \tau_\alpha K\right)\epsilon_{\partial \Sigma}\\\nonumber
&=-\frac{1}{8\pi G}\int_{\partial \Sigma} \left(-\tau^\alpha \xi^\beta \nabla_{\alpha} n_\beta+\xi^\alpha \tau_\alpha K\right)\epsilon_{\partial \Sigma}\\
&=-\frac{1}{8\pi G}\int_{\partial \Sigma}\tau^\alpha \xi^\beta\left(-K_{\alpha\beta}+\gamma_{\alpha\beta}K\right)\epsilon_{\partial \Sigma}.
\end{align}
Introducing the Brown-York stress tensor \cite{Brown:1992br}
\be\label{BYdef}
T^{\alpha\beta}\equiv \frac{2}{\sqrt{-\gamma}}\frac{\delta S}{\delta \gamma_{\alpha \beta}}=-\frac{1}{8\pi G}\left(K^{\alpha\beta}-\gamma^{\alpha\beta}K\right),
\ee
with the second equality following from \eqref{Svar} and \eqref{GRbt}, we can rewrite this as
\be\label{HxiT}
H_\xi=-\int_{\partial \Sigma}\tau^\alpha \xi^\beta T_{\alpha\beta}\epsilon_{\partial \Sigma},
\ee
which is the correct expression for the generator of a boundary isometry with killing vector $\xi^\mu$.  For fun we show how to re-derive this result using traditional non-covariant Hamiltonian methods in appendix \ref{ncapp}, where we revisit the analysis of \cite{Hawking:1995fd,Hawking:1996ww} from a slightly different point of view and extend it to obtain \eqref{HxiT} (a comparison of the lengths of the two calculations shows the advantages of the covariant formalism).

We close this section by showing how the standard ADM Hamiltonian of general relativity in asymptotically-flat spacetime \cite{arnowitt2008republication} with $d\geq 4$ can also be directly recovered from \eqref{Hresult2}.  Indeed in any asymptotically-flat spacetime we can choose coordinates $(t,x^i)$ where the metric has the form
\be
g_{\mu\nu}=\eta_{\mu\nu}+h_{\mu\nu},
\ee
where $\eta_{tt}=-1$, $\eta_{ij}=\delta_{ij}$, $h_{\mu\nu}\sim \frac{1}{r^{d-3}}$ with $r\equiv \sqrt{x^i x^i}$, and $\partial_\alpha h_{\mu\nu}\sim \frac{1}{r^{d-2}}$.  We take the spatial boundary to be at $r=r_c$ for $r_c$ some large but finite radius that we will eventually take to infinity, and we require that the pullback of $h_{\mu\nu}$ to this boundary vanish.  We discuss further the meaning of the fall-off conditions on $h_{\mu\nu}$ in section \ref{asympsec} below.  Here our goal is to compute the Hamiltonian $H_\xi$ for the vector
\be
\xi^\mu=\delta^\mu_t,
\ee
which should agree with the ADM expression.  In checking this it is sufficient to expand all quantities to linear order in $h_{\mu\nu}$, since any higher powers will give vanishing contributions to $H_\xi$ as we take $r_c \to \infty$.\footnote{In $d=4$ a term quadratic in $h_{\mu\nu}$ with no derivatives could potentially also give a non-vanishing contribution, but all terms in $Q_\xi$ and $K$ involve at least one derivative on $h_{\mu\nu}$ so this does not happen.}  Defining the ``unperturbed'' normal vector
\be
r^\mu=\delta^\mu_i x^i/r,
\ee
and using \eqref{GRQ2}, \eqref{GRXC}, and also that on $\partial \Sigma$ we have $\epsilon=-\xi\wedge r\wedge \epsilon_{\partial \Sigma}$ (see again the discussion of orientations below \eqref{stokes}), we find
\begin{align}\nonumber
Q_\xi-X_\xi\cdot C&=-\frac{1}{8\pi G}\tau^\alpha r^\beta \nabla_\alpha \xi_\beta\epsilon_{\partial \Sigma}\\
&=\frac{1}{16\pi G}\xi^\alpha\xi^\beta r^\gamma\left(2\nabla_\alpha h_{\beta\gamma}-\nabla_\gamma h_{\alpha\beta}\right)\epsilon_{\partial \Sigma}.
\end{align}
In evaluating $\left(\xi\cdot \ell\right)|_{\partial \Sigma}$ it is very useful to use the formula for $\delta K$ in \eqref{pertgeom} to compute the linear term in $h$.  Using this, and also that $\xi\cdot \epsilon_{\partial M}=-\epsilon_{\partial \Sigma}$, after some algebra we find
\begin{align}\nonumber
\left(\xi\cdot \ell\right)_{\partial \Sigma}=\frac{1}{16\pi G}\Big[&-2K_0+\delta^{ij}r^k \left(\partial_ih_{jk}-\partial_k h_{ij}\right)+\xi^\alpha\xi^\beta r^\gamma \left(\nabla_\gamma h_{\alpha\beta}-2\nabla_\alpha h_{\beta\gamma}\right)\\
&+K^{\mu\nu}\left(h_{\mu\nu}-\xi^\lambda\xi_\mu h_{\nu\lambda}+\xi_\mu\xi_\nu r^\alpha r^\beta h_{\alpha\beta}\right)+\wt{D}_\mu\left(\wt{\gamma}^{\mu\nu} r^\lambda h_{\nu\lambda}\right)\Big].
\end{align}
Here $K_0$ is the trace of the extrinsic curvature of the surface $r=r_c$ in pure Minkowski space, and $\wt{D}$ and $\wt{\gamma}_{\mu\nu}$ are the covariant derivative and induced metric on $\partial \Sigma$; the last term is thus a total derivative on $\partial \Sigma$ and does not contribute to $H_\xi$.  Moreover all terms proportional to $K^{\mu\nu}$ vanish, either because the pullback of $h_{\mu\nu}$ to the surface $r=r_c$ vanishes or because $K_{tt}=0$ in Minkowski space.  Combining these expressions we thus find that   
\be
H_\xi=\frac{1}{16\pi G}\int_{\partial \Sigma}\delta^{ij}r^k\left(\partial_ih_{jk}-\partial_k h_{ij}\right)\epsilon_{\partial \Sigma}-\frac{1}{8\pi G}\int_{\partial \Sigma}K_0\epsilon_{\partial \Sigma}.
\ee
The first term is indeed the ADM Hamiltonian, and the second is a (divergent as $r_c\to \infty$) constant on phase space.

\subsection{Brown-York stress tensor}\label{bysec}
In the previous subsection we saw that in general relativity our covariant Hamiltonian \eqref{Hresult2} was equivalent to the Brown-York expression \eqref{HxiT}.  In fact this equivalence can be extended to rather general diffeomorphism-invariant theories, as first noted in \cite{Iyer:1995kg}.  We here give an improved version of that argument, which is simpler and allows for $C\neq 0$.

Our general construction of the Hamiltonian \eqref{Hresult2} relied on choosing boundary conditions such that equation \eqref{Ceq} holds.  Here we will restrict to considering  boundary conditions where the pullback of $g_{\mu\nu}$ to $\partial M$ is fixed.  We then assume that if we allow this pullback to vary, \eqref{Ceq} is violated only as
\be\label{generalizedbc}
\Theta|_{\Gamma}+\delta \ell=dC+\frac{1}{2}T^{\mu\nu}\delta g_{\mu\nu}\epsilon_{\partial M},
\ee
where $T^{\mu\nu}$ is symmetric and obeys $T^{\mu\nu} n_\nu=0$.\footnote{In general this will require us to still impose boundary conditions on any matter fields, as well as possibly on normal derivatives of metric, and we are here assuming that a choice for these boundary conditions exists such that \eqref{generalizedbc} holds.  Moreover in \eqref{Txicalc} below we assume that any infinitesimal diffeomorphism of $\partial M$ can be extended into $M$ in a way that respects these other boundary conditions.}  We found precisely this structure in general relativity in equation \eqref{GRbt}, and in general we can think of $T^{\mu\nu}$ as the derivative of the action with respect to the boundary induced metric as in \eqref{BYdef}.  We will refer to it as the \textit{generalized Brown-York stress tensor}.

To relate $T_{\mu\nu}$ and the canonical Hamiltonian $H_\xi$, we first choose two Cauchy slices $\Sigma_-$ and $\Sigma_+$, with $\Sigma_+$ strictly in the future of $\Sigma_-$, and we then introduce a new quantity
\be
\wt{S}\equiv \int_{M_{+-}}L+\int_{\Gamma_{+-}}\ell,
\ee
where $M_{+-}$ denotes the points in $M$ which lie between $\Sigma_-$ and $\Sigma_+$ and $\Gamma_{+-}$ denoting the points in $\Gamma$ which lie between $\partial \Sigma_-$ and $\partial \Sigma_+$.  Note that we do not include any boundary terms on $\Sigma_{\pm}$.  The idea is then to compute $\delta_\xi \wt{S}$ in two different ways, where $\xi^\mu$ is an extension of an arbitrary diffeomorphism on $\partial M$ into $M$, and then to compare what we get.  The first computation uses the covariance of $L$ and $\ell$, from which we find
\be
\delta_\xi \wt{S}=\int_{\Sigma_+}\xi\cdot L-\int_{\partial \Sigma_+}\xi\cdot\ell-\int_{\Sigma_-} \xi\cdot L+\int_{\partial \Sigma_-}\xi \cdot\ell.
\ee
The signs arise from the orientation conventions explained below \eqref{stokes}.  The second computation instead uses \eqref{Lvar} and \eqref{generalizedbc}, giving
\begin{align}\nonumber
\delta_\xi \wt{S}&=\int_{\partial M_{+-}}X_\xi\cdot \Theta+\int_{\Gamma_{+-}}X_\xi\cdot \delta \ell\\
&=\int_{\Sigma_+}X_\xi\cdot \Theta-\int_{\partial \Sigma_+}X_\xi\cdot C-\int_{\Sigma_-}X_\xi\cdot \Theta+\int_{\partial\Sigma_-}X_\xi\cdot C+\int_{\Gamma_{+-}}T^{\alpha\beta}D_\alpha \xi_\beta\, \epsilon_{\partial M}.\label{Txicalc}
\end{align}
Equating these, and using \eqref{noetherdef}, \eqref{Hresult1}, and \eqref{JdQ}, we find
\begin{align}
H_\xi(\Sigma_+)-H_\xi(\Sigma_-)&=-\int_{\Gamma_{+-}}T^{\alpha\beta}D_\alpha \xi_\beta \epsilon_{\partial M}\\
&=-\int_{\partial\Sigma_+}\tau_\alpha \xi_\beta T^{\alpha\beta}\epsilon_{\partial \Sigma_+}+\int_{\partial\Sigma_-}\tau_\alpha \xi_\beta T^{\alpha\beta}\epsilon_{\partial \Sigma_-}+\int_{\Gamma_{+-}}\xi_\beta D_\alpha T^{\alpha\beta}\epsilon_{\partial M}.\label{HTresult}
\end{align}
Here all orientations are again as below \eqref{stokes}, and $\tau^\mu$ is the normal vector to $\partial \Sigma_{\pm}$ when viewed as the boundary of its past in $\partial M$.  In the first two terms on the right-hand side the minus sign in \eqref{epsilonpM} is cancelled by a minus sign arising from our orientation convention that $\partial\Gamma_{+-}=-\partial \Sigma_++\partial \Sigma_-$.  Since we can choose the restriction of $\xi^\mu$ to $\partial M$ arbitrarily, we can in particular choose it to vanish in the vicinity of $\partial \Sigma_{\pm}$ and adjust it arbitrarily elsewhere in $\Gamma_{+-}$. \eqref{HTresult} therefore then tells us that we must have $D_\alpha T^{\alpha\beta}=0$.   Moreover we can choose $\xi$ to be a Killing vector of the boundary metric in a neighborhood of $\partial \Sigma_+$, and to vanish in the vicinity of $\partial \Sigma_-$, in which case \eqref{HTresult} tells us that
\be
H_\xi(\Sigma_+)=-\int_{\partial\Sigma_+}\tau_\alpha \xi_\beta T^{\alpha\beta}\epsilon_{\partial \Sigma_+}.
\ee
We may then now take $\xi$ to be a Killing vector throughout $\partial M$, recovering \eqref{HxiT}.  Thus we see that the connection between the covariant phase space formalism and the generalized Brown-York tensor is quite close.

\subsection{Jackiw-Teitelboim gravity}\label{jtsec}
Our last example will be Jackiw-Teitelboim (JT) gravity \cite{Teitelboim:1983ux,Jackiw:1984je}, which is a simple theory of gravity coupled to a scalar in $1+1$ dimensions.  Starting with \cite{Almheiri:2014cka} it has seen considerable recent interest, in part based on its appearance within the low-temperature sector of the SYK model \cite{Jensen:2016pah,Maldacena:2016upp,Engelsoy:2016xyb}.  A covariant Hamiltonian formulation of this theory on compact space (i.e. on $\mathbb{S}^1$) was given in \cite{NavarroSalas:1992vy}, an analysis on open space (i.e. on $\mathbb{R}$) with somewhat unusual boundary conditions leading to an empty theory was given in \cite{Henneaux:1985nw}, and a Hamiltonian formulation of the theory with the ``nearly $AdS_2$'' boundary conditions appropriate for viewing it as a model of AdS/CFT was given in \cite{Harlow:2018tqv}.  In this section we describe the last case from a covariant phase space point of view.\footnote{The analysis in this section somewhat involved, we view it as a ``stress test'' of our formalism but some readers may wish to skip ahead.}

We define JT gravity to have bulk and boundary Lagrangian forms
\begin{align}\nonumber
L&=\Big(\Phi_0R+\Phi(R+2)\Big)\epsilon\\
\ell&=2\Big(\Phi_0 K+\Phi(K-1)\Big)\epsilon_{\partial M}.\label{JTL}
\end{align}
Here $\Phi$ is a dynamical scalar field, conventionally called the dilaton, $\Phi_0$ is a non-dynamical constant, and $R$ and $K$ are the intrinsic and extrinsic curvature for a dynamical metric $g_{\mu\nu}$.  Using \eqref{pertgeom}, and also that in $1+1$ dimensions we have $R_{\mu\nu}=\frac{1}{2}R g_{\mu\nu}$ and $K_{\mu\nu}=K\gamma_{\mu\nu}$, we find
\be
\delta L=E_{\Phi}\delta \Phi+E^{\mu\nu}\delta g_{\mu\nu}+d\Theta,
\ee
with
\begin{align}\nonumber
E_{\Phi}=&(R+2)\epsilon\\\nonumber
E^{\mu\nu}=&\left(\nabla^\mu \nabla^\nu\Phi-\nabla^2\Phi g^{\mu\nu}+\Phi g^{\mu\nu}\right)\epsilon\\\nonumber
\Theta=&\theta\cdot \epsilon\\\nonumber
\theta^\mu=&\left(\Phi_0+\Phi\right)\left(g^{\mu\alpha}\nabla^\beta-g^{\alpha\beta}\nabla^\mu\right)\delta g_{\alpha\beta}\\
&+\left(\nabla^\mu \Phi g^{\alpha\beta}-\nabla^\alpha \Phi g^{\mu\beta}\right)\delta g_{\alpha\beta},\label{JTtheta}
\end{align}
and also
\begin{align}
\delta\ell=\Big(&2(K-1)\delta\Phi+\left(D^\alpha \Phi n^\beta-\Phi\gamma^{\alpha\beta}\right)\delta g_{\alpha \beta}\\
&+\left(\Phi+\Phi_0\right)\left(g^{\alpha\beta} n^\lambda\nabla_\lambda \delta g_{\alpha \beta}-n^\alpha \nabla ^\beta \delta g_{\alpha \beta}\right)-D_\mu \left((\Phi_0+\Phi)\gamma^{\mu\nu}n^\alpha\delta g_{\alpha\nu}\right)\Big)\epsilon_{\partial M}.
\end{align}
Combining these we have
\be\label{JTboundaryvar}
\left(\Theta+\delta \ell\right)|_{\partial M}=\Big(2(K-1)\delta \Phi+\left(n^\mu \nabla_\mu \Phi-\Phi\right)\gamma^{\alpha \beta}\delta g_{\alpha \beta}\Big)\epsilon_{\partial M}+dC,
\ee
with
\begin{align}\nonumber
C&=c\cdot \epsilon_{\partial M}\\
c^\mu&=-(\Phi_0+\Phi)\gamma^{\mu\nu} n^\alpha\delta g_{\nu\alpha}.\label{JTC}
\end{align}
The simplest boundary conditions which respect \eqref{Ceq} are therefore those where we fix $\Phi$ and the pullback of $g_{\mu\nu}$ on $\Gamma$.  Explicitly we will take
\begin{align}\nonumber
ds^2|_{\Gamma}&=-r_c^2 dt^2\\
\Phi|_{\Gamma}&=r_c\phi_b,\label{JTbc}
\end{align}
where $\phi_b$ and $r_c$ are fixed positive constants with units of energy and length, and to recover the full $AdS_2$ geometry we take $r_c\to \infty$.  In this paper we will consider only the situation where there are two such asymptotic boundaries, as illustrated in figure \ref{jtfig}.

The Noether current for JT gravity is
\be
J_\xi = j_\xi\cdot \epsilon,
\ee
with
\begin{align}\nonumber
j_\xi^\mu=&2\nabla_\nu\left(-(\Phi_0+\Phi)\nabla^{[\mu}\xi^{\nu]}+2\nabla^{[\mu}\Phi\xi^{\nu]}\right)\\
&-2\xi_\nu\left(\nabla^\mu\nabla^\nu\Phi-g^{\mu\nu}\nabla^2\Phi+g^{\mu\nu}\Phi\right),
\end{align}
and the Noether charge is
\be\label{JTQ}
Q_\xi=-(\Phi_0+\Phi)\star d\xi+2\star\left(d\Phi\wedge \xi\right).
\ee
As in our analysis of general relativity, we can evaluate $Q_\xi$, $\xi\cdot \ell$, and $X_{\xi}\cdot C$ on $\partial \Sigma$ to compute the canonical Hamiltonian using \eqref{Hresult2}.  This again has the Brown-York form \eqref{HxiT}, with Brown-York stress tensor
\be\label{JTT}
T_{\alpha\beta}=2(n^\mu \nabla_\mu \Phi-\Phi)\gamma_{\alpha\beta}.
\ee
This can also be directly confirmed by comparing equations \eqref{generalizedbc} and \eqref{JTboundaryvar}, which is fortunate since by the argument of the previous subsection the canonical approach and the Brown-York approach must agree.

So far this analysis has paralleled that of general relativity in section \ref{grsec}, but in JT gravity with these boundary conditions one can go further and explicitly construct the phase space \cite{Harlow:2018tqv}.  We now explain how to do this using the covariant phase space formalism.  The key observation is that up to diffeomorphism all solutions of JT gravity have the form
\begin{align}\nonumber
ds^2&=-(1+x^2)d\tau^2+\frac{dx^2}{1+x^2}\\
\Phi&=\Phi_e\sqrt{1+x^2}\cos \tau,\label{JTsol}
\end{align}
where $\Phi_e$ is a dimensionless parameter that sets the value of $\Phi$ at the special point $x=\tau=0$ where the value of $\Phi$ is extremal.  Therefore the pre-phase space of JT gravity is labeled by $\Phi_e$ together with a choice of diffeomorphism.  Our task will be to clarify what part of that diffeomorphism is physical.

\bfig
\includegraphics[height=6cm]{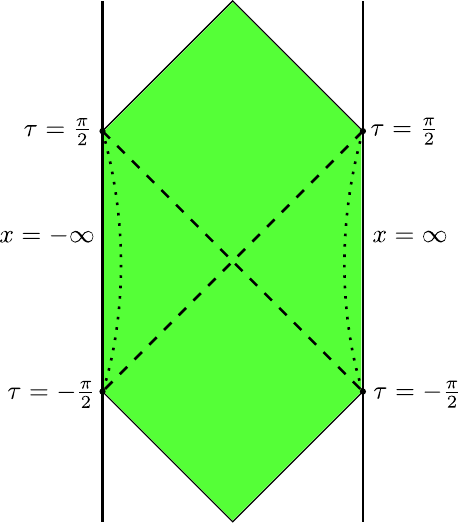}
\caption{The natural dynamical region for JT gravity: two asymptotic boundaries connected by a wormhole.  The dashed lines indicate the horizons of this wormhole, which cross at the extremal point where $\Phi=\Phi_e$. The dotted lines show where the spatial boundaries are located at finite $r_c$; these boundaries are parametrized by $t\in (-\infty,\infty)$.}\label{jtfig}
\efig
It is convenient to first say a bit more about the properties of these solutions.  The metric is just that of $AdS_2$ in global coordinates, and $x=\pm \infty$ are its two asymptotic boundaries.  In pure $AdS_2$ we would allow $\tau$ to also run from $-\infty$ to $\infty$, but here this would not respect the boundary condition \eqref{JTbc}: for $\tau$ outside of the range $(-\pi/2,\pi/2)$ the boundary value of $\Phi$ can be negative.  Therefore it is natural to consider only the dynamics of the shaded green region in figure \ref{jtfig}.  Another motivation for this is that once matter fields are included we expect the null future/past boundaries of this region to become curvature singularities, as happens in the near-extremal Reissner-Nordstr{\"o}m solution of which this is a dimensional reduction.  At finite $r_c$ we can parametrize the two asymptotic boundaries via
\begin{align}\nonumber
x_{\pm}(t)&=\pm\frac{\sqrt{ r_c^2\phi_b^2-\Phi_e^2}}{\Phi_e}\cosh \left(\frac{r_c \Phi_e}{\sqrt{ r_c^2\phi_b^2-\Phi_e^2}}\left(t+t_0^{\pm}\right)\right)\\\label{xpam}
&=\pm \frac{r_c \phi_b}{\Phi_e}\cosh \left(\frac{\Phi_e}{\phi_b}(t+t_0^{\pm})\right)+O(\frac{1}{r_c})\\\nonumber
\tan \tau_{\pm}(t)&=\sqrt{1-\frac{\Phi_e^2}{r_c^2\phi_b^2}}\sinh \left(\frac{r_c \Phi_e}{\sqrt{ r_c^2\phi_b^2-\Phi_e^2}}\left(t+t_0^{\pm}\right)\right)\\\label{taupam}
&=\sinh \left(\frac{\Phi_e}{\phi_b}(t+t_0^{\pm})\right)+O(\frac{1}{r_c^2}),
\end{align}
where $\pm$ indicate the boundaries near $x=\pm \infty$ and $t_0^{\pm}$ are arbitrary shifts of time on those boundaries.  These functions are chosen so that \eqref{JTbc} are satisfied, and one can think of $t_0^{\pm}$ as parametrizing the choice of time origin in each boundary.  In what follows the asymptotic expressions at large $r_c$ are sufficient for obtaining the final result.  We will eventually be interested in the energy of these solutions, if we consider a  diffeomorphism generator $t^\mu$ that approaches $\partial_t$ at each boundary, the Brown-York tensor \eqref{JTT} gives a Hamiltonian which evaluates (see e.g. \cite{Harlow:2018tqv}) to
\be\label{JTH}
H_t=\frac{2\Phi_e^2}{\phi_b}.
\ee

The basic technical problem we need to contend with is that in the $(\tau,x)$ coordinates the boundary locations \eqref{xpam},\eqref{taupam} depend on $\Phi_e$ and $t_{0}^\pm$, so in other words they depend on our choice of configuration and boundary Cauchy surface.  This is not consistent with our treatment of boundaries in the covariant phase space formalism, where we took the coordinate location of the boundary to be the same for all points in configuration space (and we accordingly restricted to diffeomorphisms that do not move this location).  To solve this problem we need to introduce new coordinates where the boundaries (and the Cauchy surface we use in evaluating $\wt{\Omega}$) stay put. To achieve this we first introduce a notation where we refer to the old coordinates as $x^\mu=(\tau,x)$.  We then introduce new coordinates $y^\mu=(t,y)$ related to the old ones by a diffeomorphism
\be
x^\mu=f^\mu(y),
\ee
with
\begin{align}\nonumber
f^\tau(t,y_{\pm})&=\tau_{\pm}(t)\\
f^x(t,y_{\pm}))&=x_{\pm}(t).\label{fcons}
\end{align}
In other words in the $y^\mu$ coordinates the spatial boundaries are at $y=y_{\pm}$, and on those boundaries $t$ coincides with the boundary time appearing in \eqref{JTbc}.  In these coordinates we can re-express our solutions as
\begin{align}\nonumber
g_{\mu\nu}(y)&=\partial_\mu f^\alpha \partial_\nu f^\beta g^{\{x\}}_{\alpha\beta}(f(y))\\
\Phi(y)&=\Phi^{\{x\}}(f(y)),\label{ysol}
\end{align}
where we use the superscript $\{x\}$ to indicate the specific functions appearing in \eqref{JTsol}.  We therefore can take our pre-phase space $\wt{\Pc}$ to be labeled by three real parameters $\Phi_e$, $t_0^+$, and $t_0^-$, as well as a diffeomorphism $f^\mu$ obeying \eqref{fcons}.  The crucial subtlety is that in computing variations of $\Phi$ and $g_{\mu\nu}$, we must include not only the variations of the parameters in the solutions \eqref{JTsol} (where only $\Phi_e$ appears) but also the variations of these parameters within diffeomorphisms $f^\mu$.  Once all variations have been computed, we are free to then return to the $x^\mu$ coordinates to simplify calculations.

From \eqref{JTsol} and \eqref{ysol}, the variations of the metric and dilaton in the $y$ coordinates are given by
\begin{align}\nonumber
\delta g_{\mu\nu}&=\mathcal{L}_\xi g_{\mu\nu}\\
\delta \Phi&=\Phi \frac{\delta\Phi_e}{\Phi_e}+\mathcal{L}_\xi \Phi,
\end{align}
with
\be
\xi^\mu(y)\equiv \frac{\partial (f^{-1})^\mu}{\partial x^\alpha}\Big|_{f(y)} \delta f^\alpha(y).
\ee
We emphasize that, unlike the diffeomorphisms we have considered so far, $\xi^\mu$ is a one-form on pre-phase space.
From \eqref{JTtheta}, \eqref{noetherdef}, \eqref{JdQ}, and \eqref{JTL}, we find that on pre-phase space we have
\be
\Theta|_{\wt{\Pc}}=X_\xi\cdot\Theta=J_\xi+\xi\cdot L=-2\Phi_0 \xi\cdot \epsilon+d Q_\xi.
\ee
Thus the presymplectic form is given by
\begin{align}\nonumber
\wt{\Omega}=&\delta\left[-2\Phi_0\int_\Sigma \xi\cdot \epsilon+\int_{\partial \Sigma}\left(Q_\xi-X_\xi\cdot C\right)\right]\\\nonumber
=&\delta\bigg[-2\Phi_0\int_\Sigma \xi\cdot \epsilon\\
&+2\int_{\partial \Sigma}\Big(-\tau^\mu \xi_{\mu} n^\nu \nabla_\nu \Phi+\left(\Phi_0+\Phi\right)\left(\tau^\mu \xi_\mu K-\tau^\mu \nabla_\mu\left(n_\nu \xi^\nu\right)\right)\Big)\bigg],
\end{align}
where we have used \eqref{JTQ}, \eqref{JTC}, and also that $\frac{d\Phi}{dt}|_{\Gamma}=0$. As before, $n_\mu$ is the normal form at the spatial boundary and $\tau_\mu$ is the normal form for $\partial \Sigma$ viewed as the boundary of its past in $\Gamma$.  The integral over $\partial \Sigma$ is just a sum over two points, being careful about orientation.  Computing this is a somewhat tedious exercise in working out expressions for $n_\mu$, $\tau_\mu$, and $\xi_\mu$ in the $x^\mu$ coordinates and performing the appropriate contractions.  Using the asymptotic expressions \eqref{xpam}, \eqref{taupam} for $\tau_{\pm}$ and $x_{\pm}$, we find
\begin{align}\nonumber
\xi^{\mu}n_{\mu}|_{\pm}&=-\frac{\delta \Phi_e}{\Phi_e}+O(1/r_c^2)\\
\xi^\mu \tau_\mu|_{\pm}&=\frac{r_c}{\Phi_e}\left((t+t_0^{\pm})\delta \Phi_e+ \Phi_e \delta t_0^{\pm}\right)+O(1/r_c).\label{contractions}
\end{align}
Another calculation\footnote{This calculation is simpler in the ``Schwarzschild'' coordinates
\begin{align}\nonumber
ds^2&=-(r^2-r_s^2)d\hat{t}^2+\frac{dr^2}{r^2-r_s^2}\\
\Phi&=r\phi_b,
\end{align}
with $r_s=\frac{\Phi_e}{\phi_b}$.  At finite $r_c$ the relationship between $t$ and $\hat{t}$ is $t=\sqrt{1-r_s^2/r_c^2} \hat{t}$.  These coordinates are also convenient for the calculation that gives \eqref{JTH}.}
 gives
\be
K(\Phi_0+\Phi)-n^\alpha \nabla_\alpha \Phi=\Phi_0+\frac{\Phi_e^2}{\phi_b r_c}+O(1/r_c^2).
\ee
Thus we arrive at
\be\label{almost}
\wt{\Omega}=\delta\left[-2\Phi_0 \int_\Sigma \xi\cdot \epsilon+2\sum_{\pm}\left(\Phi_0 (\xi_\mu \tau^\mu)|_\pm+\frac{\Phi_e}{\phi_b}\left((t+t_0^{\pm})\delta \Phi_e+\Phi_e \delta t_0^{\pm}\right)\right)\right],
\ee
where we have chosen our Cauchy slice $\Sigma$ to arrive at time $t$ on both boundaries.  To compute the final variation, it is useful to note that
\begin{align}\nonumber
\delta \xi^\mu&=-\frac{\partial (f^{-1})^\lambda}{\partial x^\alpha}\frac{\partial \delta f^\sigma}{\partial y^\lambda}\frac{\partial (f^{-1})^\mu}{\partial x^\sigma}\delta f^\alpha\\\nonumber
&=-\frac{\partial (f^{-1})^\mu}{\partial x^\sigma}\frac{\partial \delta f^\sigma}{\partial y^{\lambda}}\xi^\lambda\\
&=\xi^\lambda \nabla_\lambda \xi^\mu,
\end{align}
where in several places we have used the antisymmetry of $\xi^\mu \xi^\nu$ arising from the implicit wedge-product in pre-phase space.  We thus have the variation
\be
\delta(\xi\cdot \epsilon)=\nabla_\alpha\left(\xi^\alpha \xi^\mu\epsilon_{\mu\nu} dy^\nu\right)=-\frac{1}{2}\nabla_\nu\left(\xi^\alpha \xi^\beta \epsilon_{\alpha\beta}\right)dx^\nu.
\ee
where in the second equality we have used \eqref{twoformswitch}. From \eqref{contractions} and $\epsilon=\tau\wedge n$ we also have
\be
\delta\left(\tau_\mu \xi^\mu\right)=-\frac{1}{2}\xi^\alpha\xi^\beta\epsilon_{\alpha\beta},
\ee
so the terms involving $\Phi_0$ cancel in \eqref{almost}.  Finally computing the variation of the remaining boundary term, and making use of \eqref{JTH}, we have
\begin{align}\nonumber
\wt{\Omega}&=-\frac{2\Phi_e}{\phi_b}\delta(t_0^++t_0^-)\delta \Phi_e\\
&=\delta H_t\,\delta \Delta,
\end{align}
where
\be
\Delta\equiv \frac{t_0^++t_0^-}{2}.
\ee
Therefore all variations of $f^\mu$ at fixed $\Phi_e$ and $t_0^\pm$ correspond to zero modes of the pre-symplectic form, as does a variation of $t_0^+-t_0^-$ which preserves $t_0^++t_0^-$.  We thus should take the quotient of $\wt{\Pc}$ by the group action generated by these zero modes, at last obtaining a two-dimensional phase space parametrized by the energy $H_t$ and its canonical conjugate $\Delta$  \cite{Harlow:2018tqv}.\footnote{Our expression for the pre-symplectic form differs by a sign from the one given in \cite{Harlow:2018tqv}, due to a change of sign convention in its definition.}  Of course it is no surprise that the Hamiltonian is the generator of time translations, what interesting here is that it is only the combined time translation $\Delta$ which is physical, and also that there are no other degrees of freedom.  The situation is quite analogous to $1+1$-dimensional Maxwell theory on a spatial line interval, as explained in \cite{Harlow:2018tqv}.

\section{Discussion}\label{discussion}
In this final section we  consider a few interesting conceptual issues that arise in applying the covariant phase space formalism.
\subsection{Meaning of the Poisson bracket}
There is a somewhat counterintuitive property of covariant phase space.  Namely in Hamiltonian mechanics, any nonzero function on phase space generates a nontrivial evolution.  On the other hand, if we define pre-phase space as the set of solutions of the equations of motion, it seems that each point in phase space ``already knows'' its full time evolution - why do we need to evolve them at all?  And moreover doesn't this definition of phase space pick a preferred Hamiltonian?  How then are we supposed to think about evolving in this phase space using a different Hamiltonian?  We have already addressed the first question using the example \eqref{partex}: a solution which realizes some set of initial data on a Cauchy slice $\Sigma_1$ is a different solution from the one which realizes it on a distinct Cauchy slice $\Sigma_2$, and they correspond to different points in pre-phase space.  Whether or not they map to the same point in phase space is determined by whether or not there is a diffeomorphism connecting them which is generated by a zero mode of the pre-symplectic form: if there is then they coincide, while if there is not then they don't.  The second two questions are best understood by way of the \textit{Peierls bracket}, which is an old proposal for a covariant definition of the Poisson bracket \cite{Peierls:1952cb}.  We will now show that the Peierls bracket arises very naturally within the covariant phase space formalism, and thus gives an elegant interpretation of the Poisson bracket on covariant phase space.\footnote{In the absence of boundaries the relationship between these brackets has already been shown covariantly at a relatively high level of rigor in \cite{Forger:2003jm,Khavkine:2014kya}, we hope the argument given here is more digestible to a physics audience.}

In our language the insight of Peierls was to give a construction of a vector field $X_g$ on pre-phase space whose pushforward to phase space is the Hamiltonian flow vector for any $\wt{G}$-invariant function $g$ on pre-phase space (remember that $\wt{G}$ is the group whose action on pre-phase space is generated by the zero modes of $\wt{\Omega}$, usually it is the set of gauge transformations which become trivial sufficiently quickly at any boundaries).  By an analogous discussion to that around equation \eqref{inverseH}, this means a vector field such that
\be\label{Xgeq}
\delta g=-X_g\cdot \wt{\Omega}.
\ee
Given such a vector field, the Poisson bracket between $g$ and any other $\wt{G}$-invariant function $f$ is easily evaluated via
\be
\{f,g\}=X_g\cdot\delta f,
\ee
the right-hand side of which is Peierls's bracket in our notation.  The full evolution generated by $g$ may then be obtained by exponentiating this bracket.  Peierls's demonstration that his $X_g$ obeyed an equation analogous to \eqref{Xgeq} used non-covariant methods and was restricted to two-derivative theories, our goal here is to give a fully covariant demonstration which applies to arbitrary Lagrangian theories (we will however need to make a mild assumption about the initial value problem, discussed in footnote \ref{adrefn}).  

Peierls's proposal for $X_g$ is constructed as follows.  Begin with an action
\be
S_0=\int_M L+\int_{\partial M} \ell
\ee
and boundary conditions such that \eqref{Ceq} holds, and construct the associated covariant pre-phase space $\wt{\Pc}$ and phase space $\Pc$ as in section \ref{locallagsec}.  Take $g$ to be a function on configuration space whose restriction to pre-phase space is $\wt{G}$-invariant and which is constructed only using the dynamical field variables $\phi^a$ in some finite time window lying between a ``past'' Cauchy surface $\Sigma_-$ and a ``future'' Cauchy surface $\Sigma_+$.  We may then introduce a deformed action
\be
S=S_0-\lambda g,
\ee
whose equations of motion will differ from those of $S_0$ in the region $M_{+-}$ lying between $\Sigma_-$ and $\Sigma_+$.  More concretely, after enough integrations by parts we can write the variation of $g$ as
\be\label{gvar}
\delta g=\int_{M}\Delta_a^g\delta \phi^a,
\ee
where the $\Delta_a^g$ are spacetime $d$-forms that vanish outside of $M_{+-}$, and which are also functionals of the dynamical fields within $M_{+-}$.\footnote{In general these $\Delta_a^g$ will be distributional objects, involving delta-functions and so on, and may require some short-distance regularization to make precise.}  We will restrict to variations which obey the original boundary conditions for $S_0$, in which case the new action will be stationary about configurations obeying the deformed equations of motion
\be
E_a-\lambda \Delta_a^g=0.
\ee
To linear order in $\lambda$ we can write any solution of these equations as
\be
\phi^a=\phi_0^a+\lambda h^a,
\ee
where $\phi_0^a$ is a solution of the original equations of motion $E_a=0$ and $h^a$ has the property that the configuration-space vector
\be
X^{\{h\}}\equiv \int d^d x h^a(x)\frac{\delta}{\delta\phi^a(x)}
\ee
obeys
\be\label{linheq}
X^{\{h\}}\cdot \delta E_a=\Delta_a^g.
\ee
In other words $h^a$ is a solution of the linearization of the deformed equations of motion about a solution of the unperturbed equations, obeying the linearized version of the original boundary conditions.

\bfig
\includegraphics[height=4.5cm]{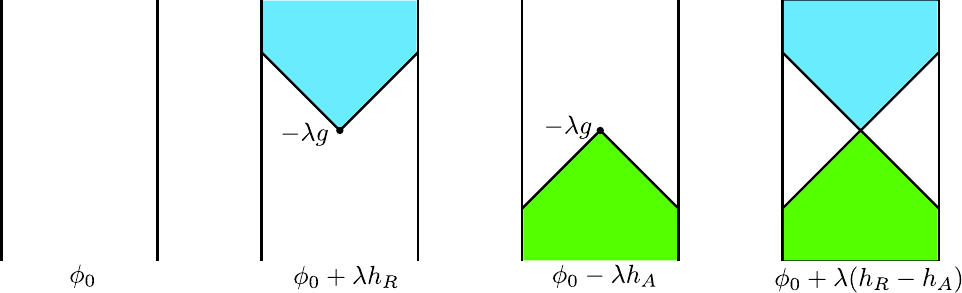}
\caption{The various solutions used in computing the Peierls bracket, in the special case where $g$ is a local operator.  The last one gives the direction in pre-phase space in which evolution by $g$ moves $\phi_0$.}\label{bracketfig}
\efig
There are two particular $h^a$ which are useful to consider: the ``advanced'' solution $h^a_A$ obeying
\be
h^a_A|_{J^+(\Sigma_+)}=0
\ee
and the ``retarded'' solution $h^a_R$ obeying
\be
h^a_R|_{J^-(\Sigma_-)}=0.
\ee
Here $J^{\pm}(\cdot)$ denotes the causal future/past of any set, so the advanced solution vanishes to the future of $\Sigma_+$ and the retarded solution vanishes to the past of $\Sigma_-$ (see figure \ref{bracketfig} for an illustration).  These two solutions are unique up to $\wt{G}$-transformations, since otherwise the difference of two distinct retarded solutions or two distinct advanced solutions would give a nontrivial solution of the unperturbed linearized equations of motion with the same initial data as the trivial solution $h^a=0$ (see \cite{Marolf:1993af}, \cite{Khavkine:2014kya} for more discussion of how gauge symmetries interact with the Peierls bracket).\footnote{We expect these advanced and retarded solutions to exist in theories with a reasonable initial-value formulation, since in such theories we do not expect that deforming the action by $-\lambda g$ to affect the set of valid initial data on Cauchy slices below $\Sigma_-$ or the set of valid final data on Cauchy slices above $\Sigma_+$.  In theories where this is not true, the Peierls bracket is not well-defined.\label{adrefn}} The proposal of Peierls is then that we should take
\be
X_g\equiv X^{\{h_R-h_A\}}=X^{\{h_R\}}-X^{\{h_A\}}.
\ee

To see that this proposal is consistent with \eqref{Xgeq}, we first note that from \eqref{linheq} we have
\be
X_g\cdot \delta E_a=(X^{\{h_R\}}-X^{\{h_A\}})\cdot \delta E_a=0.
\ee
In other words $h_R^a-h_A^a$ is a solution of the unperturbed linearized equations of motion, and we may thus interpret $X_g$ as a vector field on pre-phase space.  Now let $\Sigma$ be a Cauchy surface which is in the future of $\Sigma_+$.  We then have
\begin{align}\nonumber
-X_g\cdot \wt{\Omega}&=-X^{\{h_R\}}\cdot\int_{\Sigma}\delta(\Theta-dC)\\\nonumber
&=-X^{\{h_R\}}\cdot \int_{J^-(\Sigma)}\delta d\Theta\\\nonumber
&=-\int_{J^-(\Sigma)} X^{\{h_R\}}\cdot \delta(\delta L-E_a\delta \phi^a)\\\nonumber
&=\int_{J^-(\Sigma)}\left((X^{\{h_R\}}\cdot\delta E_a) \delta \phi^a-\delta E_a (X^{\{h_R\}}\cdot \delta \phi^a)\right)\\\nonumber
&=\int_{J^-(\Sigma)}\Delta_a^g\delta \phi^a\\
&=\delta g.
\end{align}
Here we have used that that $h_A$ has no support on $\Sigma$, that $h_R$ has no support in the distant past, that $d$ and $\delta$ commute, that $d^2=\delta^2=0$, \eqref{omegaboundary}, \eqref{Lvar}, \eqref{linheq}, \eqref{gvar}, and that the support of $\Delta_a^g$ lies in the past of $\Sigma$.  The conservation of $\wt{\Omega}$ ensures that the result actually holds for any choice of $\Sigma$.  Thus we confirm the equivalence of the Peierls and Poisson brackets, a result which in Peierls's paper was restricted to two-derivative theories and required the introduction of non-covariant methods.

In summary the Peierls bracket gives a very intuitive meaning to the Poisson bracket on covariant phase space: the quantity $\{f,g\}$ tells us the linear response of $f$ to a deformation of the action by $-g$.  The direction in pre-phase space in which $g$ evolves us is the direction of a solution of the unperturbed equations of motion obtained by starting with an unperturbed solution $\phi_0^a$ at early times, evolving forward using the deformed equations of motion to obtain a configuration at late times, and then evolving that configuration backwards using the original equations of motion.  We illustrate this in figure \ref{bracketfig}.  

\subsection{Noether's theorem}\label{noethersec}
Noether's theorem tells us that every continuous symmetry leads to a conserved charge, and in a Hamiltonian formalism any conserved charge should be the generator of a continuous symmetry.  We here show how these standard results arise within the covariant phase space formalism.  In addition to the pedagogical value of this demonstration, we will need to make use of it in the following section on asymptotic boundaries.  

We define a continuous symmetry of a Lagrangian field theory to be a vector field $X$ on the configuration space $\mathcal{C}$, which we remind the reader we define as the set of ``histories'' obeying the spatial boundary conditions but not necessarily the equations of motion,  such that
\be\label{alphadef}
X\cdot \delta L=d\alpha,
\ee
where $\alpha$ is a $d-1$ form locally constructed out of dynamical and background fields that obeys a spatial boundary condition
\be\label{alphabc}
(\alpha+X\cdot \delta\ell)|_\Gamma=d\beta,
\ee
with $\beta$ some local functional of the dynamical and background fields on $\Gamma$ and the variations of the former.\footnote{One can choose to absorb $\beta$ into a redefinition $\alpha'=\alpha-d\beta$, as we did in an earlier version of this paper to simplify calculations, but as in our discussion of $\Theta$ and $C$ doing so can lead to an $\alpha$ which is not covariant.  Keeping $\beta$ around is thus more consistent with the general philosophy of our paper \cite{Margalef-Bentabol:2020teu}.}  We emphasize that both of these equations are required to hold ``off-shell'' - they are true everywhere in configuration space.  Our goal will be to show that $X$ is tangent to pre-phase space (meaning that the flow it generates sends solutions of the equations of motion to other such solutions), that the quantity
\be\label{HXdef}
H_X\equiv \int_\Sigma\left(X\cdot \Theta-\alpha\right)+\int_{\partial \Sigma}\left(\beta-X\cdot C\right)
\ee
is conserved on pre-phase space, and that $H_X$ generates $X$ evolution in the sense that on pre-phase space
\be\label{geneq2}
\delta H_X=-X\cdot \wt{\Omega}.
\ee  
To establish the conservation of $H_X$, we note that 
\be
H_X=\int_\Sigma\left(X\cdot(\Theta-dC)-\alpha+d\beta\right),
\ee
whose integrand is a closed $d-1$ form on pre-phase space:
\begin{align}\nonumber
d(X\cdot(\Theta-dC)-\alpha)&=X\cdot d\Theta-d\alpha\\\nonumber
&=X\cdot(\delta L-E_a\delta\phi^a)-d\alpha\\
&=0.
\end{align}
Here we have used $d^2=0$, \eqref{Lvar}, \eqref{EOM}, and \eqref{alphadef}.  Moreover the pullback of the integrand to the spatial boundary $\Gamma$ vanishes by \eqref{Ceq} and \eqref{alphabc}.  Together these observations imply that indeed $H_X$ is independent of the choice of Cauchy surface $\Sigma$.  The other claimed properties of $X$ and $H_X$ are most easily derived by considering the variation of the modified action
\be
\wt{S}\equiv \int_{M_{+-}}L+\int_{\Gamma_{+-}}\ell
\ee
we introduced above in section \ref{bysec}.  Here $M_{+-}$ is the region of spacetime between a ``past'' Cauchy surface $\Sigma_-$ and a ``future'' Cauchy surface $\Sigma_+$, and $\Gamma_{+-}$ is the region of $\Gamma$ which is between $\partial \Sigma_-$ and $\partial \Sigma_+$.  The idea is to compute the Lie derivate $\mathcal{L}_X\delta \wt{S}$ in two different ways and equate them.  In the first approach we have
\be\label{Stvar1}
\mathcal{L}_X \delta \wt{S}=\int_{M_{+-}}\delta(X\cdot \delta L)+\int_{\Gamma_{+-}}\delta (X\cdot \delta \ell)=\int_{\Sigma_+}\delta (\alpha-d\beta)-\int_{\Sigma_-}\delta (\alpha-d\beta),
\ee
where we have used \eqref{alphadef} and \eqref{alphabc}.  In the second approach, we instead have
\begin{align}\nonumber
\mathcal{L}_X \delta \wt{S}=&\int_{M_{+-}}\delta(X\cdot \delta L)+\int_{\Gamma_{+-}}\delta (X\cdot \delta \ell)\\\nonumber
=&\int_{M_{+-}}\left(\delta E_a (X\cdot\delta\phi^a)+E_a \delta(X\cdot \delta \phi^a)+d\delta(X\cdot \Theta)\right)-\int_{\Gamma_{+-}}\delta(X\cdot(\Theta-dC))\\
=&\int_{\Sigma_+}(\mathcal{L}_X \Theta-d\mathcal{L}_X C)-\int_{\Sigma_-}(\mathcal{L}_X \Theta-d\mathcal{L}_X C)+\int_{M_{+-}}\Big((X\cdot \delta E_a)\delta\phi^a+E_a\mathcal{L}_X \delta \phi^a\Big),\label{Stvar2}
\end{align}
where we have used \eqref{Lvar}, \eqref{Ceq}, \eqref{cartan}, and also that
\begin{align}\nonumber
X\cdot \delta^2 L&=(X\cdot \delta E_a) \delta\phi^a-(X\cdot \delta \phi^a)\delta E_a+d(X\cdot \delta\Theta)=0\\
X\cdot \delta^2 \ell|_{\Gamma}&=\left(d(X\cdot\delta C)-X\cdot \delta \Theta\right)|_\Gamma=0.
\end{align}
Equating \eqref{Stvar1} and \eqref{Stvar2}, and using \eqref{HXdef} and \eqref{omegadef}, we find that throughout configuration space we have the one-form equation
\be
\left(\delta H_X+X\cdot \wt{\Omega}\right)\Big|_{\Sigma_+}-\left(\delta H_X+X\cdot \wt{\Omega}\right)\Big|_{\Sigma_-}=-\int_{M_{+-}}\Big((X\cdot \delta E_a)\delta\phi^a+E_a\mathcal{L}_X \delta \phi^a\Big).\label{noethereq}
\ee
We may first evaluate this equation on solutions obeying $E_a=0$ and vector fields corresponding (in the sense of equation \eqref{deltaint})  to variations about them which vanish in the neighborhood of $\Sigma_{\pm}$ but are arbitrary in the interior of $M_{+-}$, in which case we see that we must have $X\cdot \delta E_a=0$ everywhere: this shows that indeed $X$ is tangent to pre-phase space.  Therefore the right-hand side of \eqref{noethereq} vanishes for arbitrary variations in configuration space about any solution of the equations of motion.  We next consider variations in configuration space about $\wt{\Pc}$ which near $\Sigma_+$ obey the linearized equations of motion but vanish near $\Sigma_-$: we thus see that we must have \eqref{geneq2} when $H_X$ and $\wt{\Sigma}$ are evaluated on $\Sigma_+$.  Finally we observe that this statement will not be modified if we then restrict to variations which obey the linearized equations of motion everywhere, and so \eqref{geneq2} holds on pre-phase space and $H_X$ is thus indeed the generator of $X$ evolution.

Our expressions \eqref{Hresult1} and \eqref{Hresult2} for diffeomorphism generators can be viewed as a special case of this general framework, with
\begin{align}\nonumber
\alpha&=\xi\cdot L\\
\beta&=\xi\cdot \ell.
\end{align}
The reader can check that for covariant theories with $\xi$ parallel to $\Gamma$, these obey \eqref{alphadef} and \eqref{alphabc} with $X=X_\xi$, and moreover that the Hamiltonian \eqref{HXdef} is precisely the one appearing in \eqref{Hresult1}.  Indeed we could have obtained \eqref{Hresult1} entirely from this point of view to begin with, but this would have obscured the sense in which our approach is a generalization of that of \cite{Lee:1990nz,Wald:1993nt,Iyer:1994ys,Iyer:1995kg}.

\subsection{Asymptotic boundaries}\label{asympsec}
So far our general formalism has neglected the issue of the convergence of the integrals appearing in our expressions for the symplectic form and the canonical charges.  This is no issue when the Cauchy slice $\Sigma$ on which they are evaluated is a compact Riemannian manifold with boundary and all boundary conditions are finite, but in many interesting cases $\Sigma$ will either be noncompact or only be conformally compact (the latter meaning that $\Sigma$ is compact topologically but the metric and matter fields may diverge at $\partial M$).  From the point of view of this article a natural way to understand such theories is to realize them as limits of theories with an ``infrared cutoff'', as indeed we did in our discussion of the ADM energy in general relativity in section \ref{grsec} and the symplectic structure of Jackiw-Teitelboim gravity in section \ref{jtsec}.  There are however two subtleties which can arise in this procedure which we would like to discuss:
\bi
\item[(1)] If we refer to the radial location of the infrared cutoff in some coordinates as $r_c$, there can be sequences of solutions obeying the boundary conditions at finite $r_c$ which approach limits in the $r_c=\infty$ theory that have infinite energy.  These limiting solutions are those which ``have stuff all the way out'', for example in general relativity with vanishing cosmological constant we could imagine initial data where we have an infinite chain of equally-spaced copies of the Earth extending out to infinity.  Such configurations probably do not deserve the label of ``asymptotically flat'', and in any event since they have infinite energy the Hamiltonian is not well-defined on a phase space which includes them. 
\item[(2)] There may be symmetries of the $r_c=\infty$ theory which are not symmetries for any finite $r_c$.  Examples include boosts and spatial translations (and also potentially BMS transformations) of asymptotically-flat space, and also special conformal transformations in asymptotically- anti de Sitter space.  To construct the charges for these symmetries, we need to generalize our formalism to allow symmetries which are ``approximate'' at finite $r_c$.
\ei
The standard method for dealing with the first issue is to restrict to configurations in the $r_c=\infty$ theory which obey certain fall-off conditions \cite{Regge:1974zd,Ashtekar:1978zz,Henneaux:1985tv,Ashtekar:1990gc,Bartnik:2004ii}. For example in asymptotically flat space one typically restricts to metrics of the form
\be
g_{\mu\nu}=\eta_{\mu\nu}+h_{\mu\nu},
\ee
where $\eta_{\mu\nu}$ is the usual Minkowski metric in Cartesian coordinates $(t,x^i)$, and where $h_{\mu\nu}$ is required to obey\footnote{These conditions are not necessarily the full set which need to be imposed, various options are possible depending on what one is trying to achieve.  For example to get a unique set of finite Poincare generators additional ``parity conditions'' were imposed in \cite{Regge:1974zd}.  These conditions have since been relaxed in various ways, see e.g. \cite{Compere:2011ve}, leading to a larger asymptotic symmetry group that includes BMS transformations.}
\begin{align}\nonumber
h_{\mu\nu}&=O(1/r^{d-3})\\
\partial_\alpha h_{\mu\nu}&= O(1/r^{d-2}),
\end{align}
with $r\equiv \sqrt{x^i x^i}$.  These fall-off conditions do \textit{not} hold for all solutions which are limits of finite-$r_c$ configurations, and in particular imposing them ensures that the energy will be finite and thus excludes configurations with ``stuff all the way out''.  They thus must be viewed as additional requirements that are applied to the $r_c=\infty$ theory, beyond just being a limit of a sequence of finite-$r_c$ configurations obeying the boundary conditions at $r=r_c$.  This presciption may seem ad hoc, but in fact it is quite analogous to the way in which continuum quantum field theories are constructed from their lattice counterparts: as we take the lattice spacing to zero most of the states in the Hilbert space have ``too much structure at short distances'', and approach states of infinite energy in the continuum limit.  The Hilbert space of states with finite energy in the continuum is much smaller than the limit of the lattice Hilbert space, and in particular it only allows a finite number of excitations on top of the vacuum.  This resemblance is not a mere analogy, in AdS/CFT these two observations are actually dual to each other.

As for the second problem, the basic issue is that once we have an infrared regulator surface our general formalism only applies to diffeomorphisms which preserve the location of that regulator surface (such as the time translation in our discussion of the ADM energy in section \ref{grsec}).  In the limit where we remove the regulator surface, there are diffeomorphisms which would have moved it but still preserve the asymptotic fall-off conditions, and these should also be viewed as symmetries.  This phenomenon is also quite familiar from a lattice field theory point of view: introducing a short-distance regulator typically breaks many of the symmetries of a theory.  Which ones are preserved depends on the choice of regulator, but in the continuum limit they all are recovered. There are various ways that the generators for symmetries broken by the regulator can be described using our formalism, one procedure we like is the following.  Begin with a diffeomorphism generator $\xi^\mu$ which preserves the asymptotic fall-off conditions but is not parallel to the cutoff surface at $r=r_c$.  Define a flow on the regulated configuration space via
\be
\hat{X}_\xi\equiv \int d^d x \left(\mathcal{L}_\xi \phi^a +f^a\right)\frac{\delta}{\delta \phi^a(x)},
\ee
where $f^a$ is a term that ``fixes'' the violation of the boundary conditions at $r=r_c$ that is caused by applying the diffeomorphism.  In the limit that $r_c\to\infty$ we can and will take $f^a\to 0$.  At finite $r_c$ this ``corrected'' flow is not a symmetry, and in particular instead of \eqref{alphadef} we will now have
\be
\hat{X}_\xi\cdot \delta L=d\alpha+\gamma,
\ee
with $\gamma$ a $d$-form which is not necessarily exact, but which vanishes in the $r_c\to\infty$ limit at any specific point in $M$  (we can and will still take $\alpha$ to obey \eqref{alphabc}).  Our proposal is then to still define the charge for generating this approximate symmetry by equation \eqref{HXdef}.  Repeating the derivation of \eqref{noethereq} then leads to 
\be
\left(\delta H_{\hat{X}_\xi}+\hat{X}_\xi\cdot \wt{\Omega}\right)\Big|_{\Sigma_+}-\left(\delta H_{\hat{X}_\xi}+\hat{X}_\xi\cdot \wt{\Omega}\right)\Big|_{\Sigma_-}=-\int_{M_{+-}}\Big((\hat{X}_\xi\cdot \delta E_a)\delta\phi^a+E_a\mathcal{L}_{\hat{X}_\xi} \delta \phi^a-\delta \gamma\Big).
\ee
Thus if we restrict to configurations which obey fall-off conditions such that $\gamma\to 0$ in the limit that $r_c\to\infty$, we see that in the same limit $\hat{X}_\xi$ is tangent to pre-phase space and $H_{\hat{X}_\xi}$ generates $\hat{X}_\xi$ translations.  We have checked this prescription in a few simple examples, but we leave the details for future work.

\subsection{Black hole entropy}
One of the original applications of the covariant phase space formalism was in Wald's derivation of his famous entropy formula for black holes in higher-derivative gravity \cite{Wald:1993nt,Iyer:1994ys}.  This derivation is based on applying the covariant phase space formalism to a single exterior subregion of an equilibrium wormhole solution; we here show that this result is not changed by systematically including boundary terms.  Indeed let $\Sigma$ be a Cauchy surface in such a solution which contains the bifurcate horizon $\chi$, and let $\Sigma_{ext}$ be the subset of $\Sigma$ which lies between the bifurcate horizon and one of the two external spatial boundaries (we can choose either of them).  The idea is then to integrate equation \eqref{Xomega} over $\Sigma_{ext}$, with $\xi^\mu$ taken to be the Killing symmetry of the stationary black hole.  Indeed we have
\begin{align}\nonumber
-\int_{\Sigma_{ext}}X_\xi\cdot \omega=&\delta H_\xi^{ext}-\int_\chi \left(\delta Q_\xi+\delta_\xi C-\delta(X_\xi\cdot C)-\xi\cdot \Theta\right)\\\nonumber
=&\delta H_\xi^{ext}-\int_\chi \left(\delta Q_\xi+X_\xi\cdot \delta C-\xi\cdot \Theta\right)\\\nonumber
=&\delta H_\xi^{ext}-\int_\chi \delta Q_\xi\\
=&0\label{firstlaw}
\end{align}
where $H_\xi^{ext}$ denotes the contribution to $H_\xi$ from the  component of $\partial \Sigma$ which is intersected by $\Sigma_{ext}$ and we have chosen the orientation of $\chi$ so that its normal vector points towards the interior of $\Sigma_{ext}$.  In going from the first to the second line we have used \eqref{cartan}, and in going from the second to the third we have used that $\xi^\mu$ vanishes at the bifurcate horizon and also that $X_\xi$ vanishes at any point in pre-phase space where it generates a symmetry (ie where $\mathcal{L}_\xi\phi^a=0$).  The fourth line follows directly from the first as a consequence of the vanishing of $X_\xi$. Following \cite{Wald:1993nt,Iyer:1994ys} we may then interpret the equivalence of the last two lines of \eqref{firstlaw} as an expression of the first law of thermodynamics $dE=T dS$, which  leads directly to the Wald formula.

In \cite{Iyer:1994ys} the possibility of extending the Wald entropy formula to non-stationary black holes was considered, but the covariant phase space method based on the Noether charge $Q_\xi$ was dismissed on the grounds that the additive ambiguity $\Theta'=\Theta+dY$ leads to an ambiguity in $Q_\xi$ which vanishes only for stationary solutions.  We however would like to suggest that this dismissal was premature, and the issue should be reconsidered in light of the present work.  The reason is that our treatment of boundary terms actually fixes this ambiguity, leading uniquely to our $-\int_{\partial \Sigma} X_\xi \cdot C$ term in $H_\xi$.  As discussed below \eqref{Ceq}, the only remaining ambiguity is a simultaneous shift of $\Theta$ and $C$ that has no effect on $H_\xi$.  Therefore we have some hope that a generalization of the Wald formula to dynamical horizons may still be obtainable using covariant phase space techniques.  On the other hand even if the Noether charge is now unambiguous, there is no expectation of a first law for perturbations of non-stationary configurations; it is only the second law which is supposed to apply.  So it seems that some new idea (such as using the Ryu-Takayanagi formula or giving a systematic treatment of the second law) is still necessary to generalize Wald's derivation to non-stationary horizons. It would be interesting to see if our $-\int_{\partial \Sigma}X_\xi\cdot C$ term is related to the ``extrinsic curvature corrections'' appearing in the higher-derivative Ryu-Takayanagi formula of \cite{Dong:2013qoa}, and also if it might be of use in deriving a second law (see e.g. \cite{Wall:2015raa}).  To achieve this, one needs to view the exterior region as a closed dynamical system in its own right, including a careful discussion of boundary conditions on the causal horizon (knowing these will be part identifying the correct $C$ there), and it is likely that the ``edge mode'' or ``center'' degrees of freedom that arise when one defines a phase space for gravity in a subregion \cite{Donnelly:2014fua,Harlow:2015lma,Donnelly:2016auv,Speranza:2017gxd,Dong:2018seb,Kirklin:2019xug} will play an important role. In this paper we have chosen not to study null boundaries, so we leave this for future work. 

\paragraph{Acknowledgments}
We thank Bin Chen, Geoffery Comp\`{e}re, Laura Donnay, Ted Jacobson,  Jiang Long, Pujian Mao, Don Marolf, Andrea Puhm, Wei Song, Antony Speranza, Andy Strominger, and Bob Wald for useful discussions. J.q.W would like to thank the participants of the workshop ``black holes, inflation and gravitational waves'' in the Hong Kong University of Science and Technology and the workshop ``Black holes and holography workshop'' in Tsinghua Sanya International Mathematics Forum, and also the organizers of those workshops for hospitality.  D.H. would like to thank the Maryland Center for Fundamental Physics for hospitality during much of this work.  D.H. is supported by the US Department of Energy grants DE-SC0018944 and DE-SC0019127, the Simons foundation as a member of the
{\it It from Qubit} collaboration, and the MIT department of physics.  J.q.W. is supported by the Simons foundation.

\appendix
\section{Non-covariant Hamiltonian analysis of general relativity}\label{ncapp}
In this appendix we show how to obtain the canonical Hamiltonian \eqref{HxiT} for any boundary-preserving diffeomorphism generator $\xi^\mu$ in general relativity directly from the traditional non-covariant approach.  The idea of such a calculation goes back quite a ways \cite{arnowitt2008republication,DeWitt:1967yk,Regge:1974zd}, but it was not until \cite{Hawking:1995fd,Hawking:1996ww} that a systematic treatment of boundary terms starting from the action formalism was given.  In that treatment the resulting Hamiltonian was presented in a somewhat unusual form.  In this appendix we redo that analysis in a somewhat different manner, in particular we are able to allow the Cauchy slice to intersect the spatial boundary non-orthogonally without needing to discuss Hayward terms \cite{Hayward:1993my}, and our presentation results in the standard Brown-York expression \eqref{HxiT} for the Hamiltonian.  This calculation is not necessary for the logical flow of our paper, but we find it useful to illustrate the relative convenience of the covariant formalism by comparison.

We begin by choosing a set of Cauchy surfaces $\Sigma_t$ which foliate our (globally-hyperbolic) spacetime $M$ and are labelled by a time coordinate $t$.  We also (nonuniquely) choose coordinates on each slice such that we can view the spacetime as $\mathbb{R}\times \Sigma$, with $\Sigma$ some $d-1$ manifold which is homeomorphic to each $\Sigma_t$. We are interested in finding the Hamiltonian $H_\xi$ for the diffeomorphisms generated by the vector field
\be
\xi^\mu=\delta^\mu_t,
\ee
which we can represent using the ADM decomposition\footnote{Note that in the covariant version of the ADM formalism we use here, the quantities $N^\mu$ and $\hat{n}^\mu$ are vectors on the full spacetime, indices are raised and lowered using the full spacetime metric $g_{\mu\nu}$, and $\nabla_\mu$ is the ordinary covariant derivative.  All notation for normal forms and extrinsic curvatures is as introduced around \eqref{normaldef}-\eqref{Kdef}.}
\be\label{admdc}
\xi^\mu=-N \hat{n}^\mu+N^\mu.
\ee
Here $\hat{n}_\mu$ is the normal form to the Cauchy slices $\Sigma_t$ (not to be confused with $n_\mu$ the normal form to the boundary $\partial M$), and we require that $N>0$ and $N^\mu\hat{n}_\mu=0$.
Explicitly,
\be
\hat{n}_\mu=N \delta_\mu^t.
\ee
$N$ is called the \textit{lapse}: it measures how fast proper time elapses on a geodesic normal to $\Sigma_t$ as we change $t$. $N^\mu$ is called the \textit{shift}: it measures how much the coordinates we've chosen on the $\Sigma_t$ shift as we change $t$ relative to what we would have gotten by connecting them using normal geodesics.

We now study general relativity with the action given by \eqref{GRact}.  The basic idea is to view the induced metric $\hat{\gamma}_{\mu\nu}\equiv g_{\mu\nu}+\hat{n}_\mu \hat{n}_\nu$ on each Cauchy slice $\Sigma_t$ as the ``position'' degrees of freedom, identify their conjugate canonical momenta, and then compute the Hamiltonian via the usual formula $H=p \dot{q}-L$.  We therefore need to re-express the action in a way that makes manifest all occurrences of the time derivative
\be
\dot{\hat{\gamma}}_{\mu\nu}\equiv \hat{\gamma}_\mu^{\phantom{\mu}\alpha}\hat{\gamma}_\nu^{\phantom{\nu}\beta}\mathcal{L}_\xi \hat{\gamma}_{\alpha\beta}.
\ee
This is facilitated by the Gauss-Codacci equation
\be\label{gceq}
R=\hat{R}+\hat{K}_{\mu\nu}\hat{K}^{\mu\nu}-\hat{K}^2+2\nabla_\mu\left(\hat{n}^\mu\nabla_\nu \hat{n}^\nu\right)-2\nabla_\mu\left(\hat{n}^\nu\nabla_\nu \hat{n}^\mu\right),
\ee
where $\hat{K}_{\mu\nu}\equiv \hat{\gamma}_\mu^{\phantom{\mu}\lambda}\nabla_\lambda \hat{n}_\nu$  is the extrinsic curvature of the $\Sigma_t$ in $M$ (not to be confused with $K_{\mu\nu}$ the extrinsic curvature at the boundary $\partial M$) and $\hat{R}$ is the Ricci scalar for the induced metric on the $\Sigma_t$ \cite{Wald:1984rg}.  Using \eqref{gceq} on \eqref{GRact} we have
\be
S=\frac{1}{16\pi G}\left[\int_M\left(\hat{R}+\hat{K}_{\mu\nu}\hat{K}^{\mu\nu}-\hat{K}^2-2\Lambda\right)\epsilon+2\int_{\partial M}\left(n_\mu \hat{n}^\mu\nabla_\nu \hat{n}^\nu-n_\mu\hat{n}^\nu \nabla_\nu \hat{n}^\mu+K\right)\epsilon_{\partial M}\right].
\ee
The only time derivatives in the non-boundary part of this action arise from the extrinsic curvatures via
\be
\hat{K}_{\mu\nu}=\frac{1}{2N}\left(\hat{D}_\mu N_\nu+\hat{D}_\nu N_\mu-\dot{\hat{\gamma}}_{\mu\nu}\right),
\ee
where $\hat{D}$ is the hypersurface covariant derivative on $\Sigma_t$, defined as in \eqref{Ddef}.  Introducing the normal form $\tau_\mu$ to $\partial \Sigma$ within $\partial M$, from $\epsilon=\hat{n}\wedge \epsilon_\Sigma$ and \eqref{epsilonpM} we have
\begin{align}\nonumber
\int_M \epsilon&=\int dt \int_{\Sigma} N \epsilon_{\Sigma}\\
\int_{\partial M}\epsilon_{\partial M}&=\int dt \int_{\partial\Sigma} \tau_\mu \xi^\mu \epsilon_{\partial \Sigma},
\end{align}
which we may then use to rewrite the action (after discarding terms at the future/past boundaries) as
\be
S=\int dt \hat{L},
\ee
with
\begin{align}\nonumber
\hat{L}=\frac{1}{16\pi G}\Big[&\int_\Sigma N \left(\hat{R}+\hat{K}_{\mu\nu}\hat{K}^{\mu\nu}-\hat{K}^2-2\Lambda\right)\epsilon_\Sigma\\
&+2\int_{\partial \Sigma}\tau_\rho \xi^\rho\left(n_\mu \hat{n}^\mu\nabla_\nu \hat{n}^\nu-n_\mu\hat{n}^\nu \nabla_\nu \hat{n}^\mu+K\right)\epsilon_{\partial \Sigma}\Big].\label{noncact}
\end{align}
At this point the authors of \cite{Hawking:1995fd} chose to set $\hat{n}_\mu n^\mu=0$, in which case the boundary term in \eqref{noncact} just becomes the integral of the extrinsic curvature of $\partial \Sigma$ within $\Sigma$, which manifestly depends only on the induced metric on $\Sigma_t$ and not its time derivative.  We however will not assume this (dropping this assumption was also the goal of \cite{Hawking:1996ww}), and will instead observe that a short calculation shows that
\be
n_\mu \hat{n}^\mu\nabla_\nu \hat{n}^\nu-n_\mu\hat{n}^\nu \nabla_\nu \hat{n}^\mu+K=\hat{D}_\mu \left(\hat{\gamma}^{\mu\nu}n_\nu\right)+\left(\xi^\rho-N^\rho\right)N^{-1}\nabla_\rho\left(n_\mu\hat{n}^\mu\right).
\ee
Thus in addition to the time derivatives of $\hat{\gamma}_{\mu\nu}$ arising from the extrinsic curvatures, in the boundary term there is also a time-derivative of the quantity $n_\mu \hat{n}^\mu$, which we therefore must view as an additional dynamical degree of freedom.  The canonical momenta conjugate to $\hat{\gamma}_{\mu\nu}$ and $n_\mu \hat{n}^\mu$ which follow from \eqref{noncact} are
\begin{align}\nonumber
P^{\mu\nu}&=-\frac{\sqrt{\hat{\gamma}}}{16\pi G}\left(\hat{K}^{\mu\nu}-\hat{\gamma}^{\mu\nu} \hat{K}\right)\\
p&=\frac{1}{8\pi G}\frac{\tau_\mu \xi^\mu}{N}\sqrt{\gamma_{\partial \Sigma}},
\end{align}
where $\gamma_{\partial \Sigma}$ is the determinant of the induced metric on $\partial \Sigma$.  We thus may compute the Hamiltonian via
\be
H_\xi = \int_\Sigma d^{d-1}x P^{\mu\nu}\dot{\hat{\gamma}}_{\mu\nu}+\int_{\partial \Sigma}d^{d-2}x \,p\,\xi^\mu \nabla_\mu\left(n_\nu \hat{n}^\nu\right)-\hat{L}.
\ee
Substituting the above formulas and doing a bit of algebra, we find
\begin{align}\nonumber
H_\xi=&\int_{\Sigma} \left[-2N_\mu \frac{1}{\sqrt{\hat{\gamma}}}\hat{D}_\nu P^{\mu\nu}+\frac{N}{16\pi G}\left(-\hat{R}+2\Lambda+\frac{(16\pi G)^2}{\hat{\gamma}}\left(P_{\mu\nu}P^{\mu\nu}-\frac{1}{d-2}P^2\right)\right)\right]\epsilon_{\Sigma}\\
&+\int_{\partial \Sigma}\left[\frac{2}{\sqrt{\hat{\gamma}}}P^{\mu\nu}r_\mu N_\nu +\frac{\tau_\mu \xi^\mu}{8\pi G}\left(N^{-1}\xi^\mu \nabla_\mu\left(n_\nu \hat{n}^\nu\right)-n_\mu \hat{n}^\mu\nabla_\nu \hat{n}^\nu+n_\mu\hat{n}^\nu \nabla_\nu \hat{n}^\mu-K\right)\right]\epsilon_{\partial \Sigma}\label{bigHeq}
\end{align}
In the second line the quantity $r_\mu$ is the normal form to $\partial \Sigma$ within $\Sigma$.  $r_\mu$ and $\tau_\mu$ are related to $n_\mu$ and $\hat{n}_\mu$ via
\begin{align}\nonumber
r_\mu&=\alpha \hat{\gamma}_\mu^{\phantom{\mu}\nu}n_\nu\\
\tau_\mu&=\alpha \gamma_\mu^{\phantom{\mu}\nu}\hat{n}_\nu,\label{rtaueq}
\end{align}
with
\be
\alpha=\frac{1}{\sqrt{1+(n_\mu \hat{n}^\mu)^2}}.
\ee
The terms multiplying $N_\mu$ and $N$ in the first line of \eqref{bigHeq} are just the shift and Hamiltonian constraint equations of general relativity, which vanish on shell, so as expected the on-shell Hamiltonian is a pure boundary term (the second line of \eqref{bigHeq}).  Moreover we can simplify the second line using the definitions of $P^{\mu\nu}$ and $\hat{K}$, and also \eqref{rtaueq} and \eqref{admdc}, to find that indeed
\be
H_\xi=\frac{1}{8\pi G}\int_{\partial \Sigma}\tau^\alpha\xi^\beta \left(K_{\alpha \beta}-\gamma_{\alpha\beta}K\right),
\ee
as needed to match \eqref{HxiT}.
\bibliographystyle{jhep}
\bibliography{bibliography}

\providecommand{\href}[2]{#2}\begingroup\raggedright\begin{thebibliography}{10}

\bibitem{Witten:1986qs}
E.~Witten, {\it {Interacting Field Theory of Open Superstrings}},  {\em Nucl.
  Phys.} {\bf B276} (1986) 291--324.

\bibitem{zuckerman1987action}
G.~J. Zuckerman, {\it Action principles and global geometry},  in {\em
  Mathematical Aspects of String Theory}, pp.~259--284.
\newblock World Scientific, 1987.

\bibitem{crnkovic1987covariant}
C.~Crnkovic and E.~Witten, {\it Covariant description of canonical formalism in
  geometrical theories},  in {\em Three hundred years of gravitation},
  pp.~676--684.
\newblock Cambridge University Press, 1987.

\bibitem{Crnkovic:1987tz}
C.~Crnkovic, {\it {Symplectic Geometry of the Covariant Phase Space,
  Superstrings and Superspace}},  {\em Class. Quant. Grav.} {\bf 5} (1988)
  1557--1575.

\bibitem{Lee:1990nz}
J.~Lee and R.~M. Wald, {\it {Local symmetries and constraints}},  {\em J. Math.
  Phys.} {\bf 31} (1990) 725--743.

\bibitem{Wald:1993nt}
R.~M. Wald, {\it {Black hole entropy is the Noether charge}},  {\em Phys. Rev.}
  {\bf D48} (1993), no.~8 R3427--R3431,
  [\href{http://arxiv.org/abs/gr-qc/9307038}{{\tt gr-qc/9307038}}].

\bibitem{Iyer:1994ys}
V.~Iyer and R.~M. Wald, {\it {Some properties of Noether charge and a proposal
  for dynamical black hole entropy}},  {\em Phys. Rev.} {\bf D50} (1994)
  846--864, [\href{http://arxiv.org/abs/gr-qc/9403028}{{\tt gr-qc/9403028}}].

\bibitem{Iyer:1995kg}
V.~Iyer and R.~M. Wald, {\it {A Comparison of Noether charge and Euclidean
  methods for computing the entropy of stationary black holes}},  {\em Phys.
  Rev.} {\bf D52} (1995) 4430--4439,
  [\href{http://arxiv.org/abs/gr-qc/9503052}{{\tt gr-qc/9503052}}].

\bibitem{Wald:1999wa}
R.~M. Wald and A.~Zoupas, {\it {A General definition of 'conserved quantities'
  in general relativity and other theories of gravity}},  {\em Phys. Rev.} {\bf
  D61} (2000) 084027, [\href{http://arxiv.org/abs/gr-qc/9911095}{{\tt
  gr-qc/9911095}}].

\bibitem{Khavkine:2014kya}
I.~Khavkine, {\it {Covariant phase space, constraints, gauge and the Peierls
  formula}},  {\em Int. J. Mod. Phys.} {\bf A29} (2014), no.~5 1430009,
  [\href{http://arxiv.org/abs/1402.1282}{{\tt arXiv:1402.1282}}].

\bibitem{Hollands:2005wt}
S.~Hollands, A.~Ishibashi, and D.~Marolf, {\it {Comparison between various
  notions of conserved charges in asymptotically AdS-spacetimes}},  {\em Class.
  Quant. Grav.} {\bf 22} (2005) 2881--2920,
  [\href{http://arxiv.org/abs/hep-th/0503045}{{\tt hep-th/0503045}}].

\bibitem{Compere:2008us}
G.~Compere and D.~Marolf, {\it {Setting the boundary free in AdS/CFT}},  {\em
  Class. Quant. Grav.} {\bf 25} (2008) 195014,
  [\href{http://arxiv.org/abs/0805.1902}{{\tt arXiv:0805.1902}}].

\bibitem{Faulkner:2013ica}
T.~Faulkner, M.~Guica, T.~Hartman, R.~C. Myers, and M.~Van~Raamsdonk, {\it
  {Gravitation from Entanglement in Holographic CFTs}},  {\em JHEP} {\bf 03}
  (2014) 051, [\href{http://arxiv.org/abs/1312.7856}{{\tt arXiv:1312.7856}}].

\bibitem{Andrade:2015gja}
T.~Andrade, W.~R. Kelly, D.~Marolf, and J.~E. Santos, {\it {On the stability of
  gravity with Dirichlet walls}},  {\em Class. Quant. Grav.} {\bf 32} (2015),
  no.~23 235006, [\href{http://arxiv.org/abs/1504.07580}{{\tt
  arXiv:1504.07580}}].

\bibitem{Jafferis:2015del}
D.~L. Jafferis, A.~Lewkowycz, J.~Maldacena, and S.~J. Suh, {\it {Relative
  entropy equals bulk relative entropy}},  {\em JHEP} {\bf 06} (2016) 004,
  [\href{http://arxiv.org/abs/1512.06431}{{\tt arXiv:1512.06431}}].

\bibitem{Lashkari:2016idm}
N.~Lashkari, J.~Lin, H.~Ooguri, B.~Stoica, and M.~Van~Raamsdonk, {\it
  {Gravitational positive energy theorems from information inequalities}},
  {\em PTEP} {\bf 2016} (2016), no.~12 12C109,
  [\href{http://arxiv.org/abs/1605.01075}{{\tt arXiv:1605.01075}}].

\bibitem{Dong:2018seb}
X.~Dong, D.~Harlow, and D.~Marolf, {\it {Flat entanglement spectra in
  fixed-area states of quantum gravity}},
  \href{http://arxiv.org/abs/1811.05382}{{\tt arXiv:1811.05382}}.

\bibitem{Belin:2018fxe}
A.~Belin, A.~Lewkowycz, and G.~Sárosi, {\it {The boundary dual of the bulk
  symplectic form}},  {\em Phys. Lett.} {\bf B789} (2019) 71--75,
  [\href{http://arxiv.org/abs/1806.10144}{{\tt arXiv:1806.10144}}].

\bibitem{Belin:2018bpg}
A.~Belin, A.~Lewkowycz, and G.~Sárosi, {\it {Complexity and the bulk volume, a
  new York time story}},  \href{http://arxiv.org/abs/1811.03097}{{\tt
  arXiv:1811.03097}}.

\bibitem{Compere:2011ve}
G.~Compere and F.~Dehouck, {\it {Relaxing the Parity Conditions of
  Asymptotically Flat Gravity}},  {\em Class. Quant. Grav.} {\bf 28} (2011)
  245016, [\href{http://arxiv.org/abs/1106.4045}{{\tt arXiv:1106.4045}}].
  [Erratum: Class. Quant. Grav.30,039501(2013)].

\bibitem{Chandrasekaran:2018aop}
V.~Chandrasekaran, a.~. Flanagan, and K.~Prabhu, {\it {Symmetries and charges
  of general relativity at null boundaries}},  {\em JHEP} {\bf 11} (2018) 125,
  [\href{http://arxiv.org/abs/1807.11499}{{\tt arXiv:1807.11499}}].

\bibitem{Carlip:1999cy}
S.~Carlip, {\it {Entropy from conformal field theory at Killing horizons}},
  {\em Class. Quant. Grav.} {\bf 16} (1999) 3327--3348,
  [\href{http://arxiv.org/abs/gr-qc/9906126}{{\tt gr-qc/9906126}}].

\bibitem{Haco:2018ske}
S.~Haco, S.~W. Hawking, M.~J. Perry, and A.~Strominger, {\it {Black Hole
  Entropy and Soft Hair}},  {\em JHEP} {\bf 12} (2018) 098,
  [\href{http://arxiv.org/abs/1810.01847}{{\tt arXiv:1810.01847}}].

\bibitem{Haco:2019ggi}
S.~Haco, M.~J. Perry, and A.~Strominger, {\it {Kerr-Newman Black Hole Entropy
  and Soft Hair}},  \href{http://arxiv.org/abs/1902.02247}{{\tt
  arXiv:1902.02247}}.

\bibitem{Regge:1974zd}
T.~Regge and C.~Teitelboim, {\it {Role of Surface Integrals in the Hamiltonian
  Formulation of General Relativity}},  {\em Annals Phys.} {\bf 88} (1974) 286.

\bibitem{Hawking:1995fd}
S.~W. Hawking and G.~T. Horowitz, {\it {The Gravitational Hamiltonian, action,
  entropy and surface terms}},  {\em Class. Quant. Grav.} {\bf 13} (1996)
  1487--1498, [\href{http://arxiv.org/abs/gr-qc/9501014}{{\tt gr-qc/9501014}}].

\bibitem{Hawking:1996ww}
S.~Hawking and C.~Hunter, {\it {The Gravitational Hamiltonian in the presence
  of nonorthogonal boundaries}},  {\em Class. Quant. Grav.} {\bf 13} (1996)
  2735--2752, [\href{http://arxiv.org/abs/gr-qc/9603050}{{\tt gr-qc/9603050}}].

\bibitem{wald1990identically}
R.~M. Wald, {\it On identically closed forms locally constructed from a field},
   {\em Journal of mathematical physics} {\bf 31} (1990), no.~10 2378--2384.

\bibitem{Julia:2002df}
B.~Julia and S.~Silva, {\it {On covariant phase space methods}},
  \href{http://arxiv.org/abs/hep-th/0205072}{{\tt hep-th/0205072}}.

\bibitem{Papadimitriou:2005ii}
I.~Papadimitriou and K.~Skenderis, {\it {Thermodynamics of asymptotically
  locally AdS spacetimes}},  {\em JHEP} {\bf 08} (2005) 004,
  [\href{http://arxiv.org/abs/hep-th/0505190}{{\tt hep-th/0505190}}].

\bibitem{Andrade:2015fna}
T.~Andrade and D.~Marolf, {\it {Asymptotic Symmetries from finite boxes}},
  {\em Class. Quant. Grav.} {\bf 33} (2016), no.~1 015013,
  [\href{http://arxiv.org/abs/1508.02515}{{\tt arXiv:1508.02515}}].

\bibitem{Donnelly:2016rvo}
W.~Donnelly and S.~B. Giddings, {\it {Observables, gravitational dressing, and
  obstructions to locality and subsystems}},  {\em Phys. Rev.} {\bf D94}
  (2016), no.~10 104038, [\href{http://arxiv.org/abs/1607.01025}{{\tt
  arXiv:1607.01025}}].

\bibitem{Giddings:2018umg}
S.~B. Giddings and A.~Kinsella, {\it {Gauge-invariant observables,
  gravitational dressings, and holography in AdS}},  {\em JHEP} {\bf 11} (2018)
  074, [\href{http://arxiv.org/abs/1802.01602}{{\tt arXiv:1802.01602}}].

\bibitem{Barnich:2001jy}
G.~Barnich and F.~Brandt, {\it {Covariant theory of asymptotic symmetries,
  conservation laws and central charges}},  {\em Nucl. Phys.} {\bf B633} (2002)
  3--82, [\href{http://arxiv.org/abs/hep-th/0111246}{{\tt hep-th/0111246}}].

\bibitem{Barnich:2007bf}
G.~Barnich and G.~Compere, {\it {Surface charge algebra in gauge theories and
  thermodynamic integrability}},  {\em J. Math. Phys.} {\bf 49} (2008) 042901,
  [\href{http://arxiv.org/abs/0708.2378}{{\tt arXiv:0708.2378}}].

\bibitem{almaraz2014positive}
S.~Almaraz, E.~Barbosa, and L.~L. de~Lima, {\it A positive mass theorem for
  asymptotically flat manifolds with a non-compact boundary},  {\em arXiv
  preprint arXiv:1407.0673} (2014).

\bibitem{almaraz2018mass}
S.~Almaraz and L.~L. de~Lima, {\it The mass of an asymptotically hyperbolic
  manifold with a noncompact boundary},  {\em arXiv preprint arXiv:1811.06913}
  (2018).

\bibitem{almaraz2019spacetime}
S.~Almaraz, L.~L. de~Lima, and L.~Mari, {\it Spacetime positive mass theorems
  for initial data sets with noncompact boundary},  {\em arXiv preprint
  arXiv:1907.02023} (2019).

\bibitem{Prabhu:2015vua}
K.~Prabhu, {\it {The First Law of Black Hole Mechanics for Fields with Internal
  Gauge Freedom}},  {\em Class. Quant. Grav.} {\bf 34} (2017), no.~3 035011,
  [\href{http://arxiv.org/abs/1511.00388}{{\tt arXiv:1511.00388}}].

\bibitem{arnold2007mathematical}
V.~I. Arnold, V.~V. Kozlov, and A.~I. Neishtadt, {\em Mathematical aspects of
  classical and celestial mechanics}, vol.~3.
\newblock Springer Science \& Business Media, 2007.

\bibitem{Noether:1918zz}
E.~Noether, {\it {Invariant Variation Problems}},  {\em Gott. Nachr.} {\bf
  1918} (1918) 235--257, [\href{http://arxiv.org/abs/physics/0503066}{{\tt
  physics/0503066}}]. [Transp. Theory Statist. Phys.1,186(1971)].

\bibitem{lang2012fundamentals}
S.~Lang, {\em Fundamentals of differential geometry}, vol.~191.
\newblock Springer Science \& Business Media, 2012.

\bibitem{marsden1981lectures}
J.~E. Marsden, {\em Lectures on geometric methods in mathematical physics}.
\newblock SIAM, 1981.

\bibitem{abraham1978foundations}
R.~Abraham, J.~E. Marsden, and J.~E. Marsden, {\em Foundations of mechanics},
  vol.~36.
\newblock Benjamin/Cummings Publishing Company Reading, Massachusetts, 1978.

\bibitem{Wald:1984rg}
R.~M. Wald, {\em {General Relativity}}.
\newblock Chicago Univ. Pr., Chicago, USA, 1984.

\bibitem{Carroll:2004st}
S.~M. Carroll, {\em {Spacetime and geometry: An introduction to general
  relativity}}.
\newblock 2004.

\bibitem{Brown:1992br}
J.~D. Brown and J.~W. York, Jr., {\it {Quasilocal energy and conserved charges
  derived from the gravitational action}},  {\em Phys. Rev.} {\bf D47} (1993)
  1407--1419, [\href{http://arxiv.org/abs/gr-qc/9209012}{{\tt gr-qc/9209012}}].

\bibitem{arnowitt2008republication}
R.~Arnowitt, S.~Deser, and C.~W. Misner, {\it Republication of: The dynamics of
  general relativity},  {\em General Relativity and Gravitation} {\bf 40}
  (2008), no.~9 1997--2027.

\bibitem{Teitelboim:1983ux}
C.~Teitelboim, {\it {Gravitation and Hamiltonian Structure in Two Space-Time
  Dimensions}},  {\em Phys. Lett.} {\bf 126B} (1983) 41--45.

\bibitem{Jackiw:1984je}
R.~Jackiw, {\it {Lower Dimensional Gravity}},  {\em Nucl. Phys.} {\bf B252}
  (1985) 343--356.

\bibitem{Almheiri:2014cka}
A.~Almheiri and J.~Polchinski, {\it {Models of AdS$_{2}$ backreaction and
  holography}},  {\em JHEP} {\bf 11} (2015) 014,
  [\href{http://arxiv.org/abs/1402.6334}{{\tt arXiv:1402.6334}}].

\bibitem{Jensen:2016pah}
K.~Jensen, {\it {Chaos in AdS$_2$ Holography}},  {\em Phys. Rev. Lett.} {\bf
  117} (2016), no.~11 111601, [\href{http://arxiv.org/abs/1605.06098}{{\tt
  arXiv:1605.06098}}].

\bibitem{Maldacena:2016upp}
J.~Maldacena, D.~Stanford, and Z.~Yang, {\it {Conformal symmetry and its
  breaking in two dimensional Nearly Anti-de-Sitter space}},  {\em PTEP} {\bf
  2016} (2016), no.~12 12C104, [\href{http://arxiv.org/abs/1606.01857}{{\tt
  arXiv:1606.01857}}].

\bibitem{Engelsoy:2016xyb}
J.~Engels{\"o}y, T.~G. Mertens, and H.~Verlinde, {\it {An investigation of
  AdS$_{2}$ backreaction and holography}},  {\em JHEP} {\bf 07} (2016) 139,
  [\href{http://arxiv.org/abs/1606.03438}{{\tt arXiv:1606.03438}}].

\bibitem{NavarroSalas:1992vy}
J.~Navarro-Salas, M.~Navarro, and V.~Aldaya, {\it {Covariant phase space
  quantization of the Jackiw-Teitelboim model of 2-D gravity}},  {\em Phys.
  Lett.} {\bf B292} (1992) 19--24.

\bibitem{Henneaux:1985nw}
M.~Henneaux, {\it {QUANTUM GRAVITY IN TWO-DIMENSIONS: EXACT SOLUTION OF THE
  JACKIW MODEL}},  {\em Phys. Rev. Lett.} {\bf 54} (1985) 959--962.

\bibitem{Harlow:2018tqv}
D.~Harlow and D.~Jafferis, {\it {The Factorization Problem in Jackiw-Teitelboim
  Gravity}},  \href{http://arxiv.org/abs/1804.01081}{{\tt arXiv:1804.01081}}.

\bibitem{Peierls:1952cb}
R.~E. Peierls, {\it {The Commutation laws of relativistic field theory}},  {\em
  Proc. Roy. Soc. Lond.} {\bf A214} (1952) 143--157.

\bibitem{Forger:2003jm}
M.~Forger and S.~V. Romero, {\it {Covariant poisson brackets in geometric field
  theory}},  {\em Commun. Math. Phys.} {\bf 256} (2005) 375--410,
  [\href{http://arxiv.org/abs/math-ph/0408008}{{\tt math-ph/0408008}}].

\bibitem{Marolf:1993af}
D.~M. Marolf, {\it {The Generalized Peierls bracket}},  {\em Annals Phys.} {\bf
  236} (1994) 392--412, [\href{http://arxiv.org/abs/hep-th/9308150}{{\tt
  hep-th/9308150}}].

\bibitem{Margalef-Bentabol:2020teu}
J.~Margalef-Bentabol and E.~J. Villase\~{n}or, {\it {Geometric formulation of
  the Covariant Phase Space methods with boundaries}},
  \href{http://arxiv.org/abs/2008.01842}{{\tt arXiv:2008.01842}}.

\bibitem{Ashtekar:1978zz}
A.~Ashtekar and R.~O. Hansen, {\it {A unified treatment of null and spatial
  infinity in general relativity. I - Universal structure, asymptotic
  symmetries, and conserved quantities at spatial infinity}},  {\em J. Math.
  Phys.} {\bf 19} (1978) 1542--1566.

\bibitem{Henneaux:1985tv}
M.~Henneaux and C.~Teitelboim, {\it {Asymptotically anti-De Sitter Spaces}},
  {\em Commun. Math. Phys.} {\bf 98} (1985) 391--424.

\bibitem{Ashtekar:1990gc}
A.~Ashtekar, L.~Bombelli, and O.~Reula, {\it The covariant phase space of
  asymptotically flat gravitational fields},  in {\em Analysis, Geometry, and
  Mechanics: 200 Years After Lagrange} (M.~Francaviglia and D.~Holm, eds.).
\newblock North Holland, Amsterdam, 1990.

\bibitem{Bartnik:2004ii}
R.~Bartnik, {\it {Phase space for the Einstein equations}},  {\em Submitted to:
  Commun. Anal. Geom.} (2004) [\href{http://arxiv.org/abs/gr-qc/0402070}{{\tt
  gr-qc/0402070}}].

\bibitem{Dong:2013qoa}
X.~Dong, {\it {Holographic Entanglement Entropy for General Higher Derivative
  Gravity}},  {\em JHEP} {\bf 01} (2014) 044,
  [\href{http://arxiv.org/abs/1310.5713}{{\tt arXiv:1310.5713}}].

\bibitem{Wall:2015raa}
A.~C. Wall, {\it {A Second Law for Higher Curvature Gravity}},  {\em Int. J.
  Mod. Phys.} {\bf D24} (2015), no.~12 1544014,
  [\href{http://arxiv.org/abs/1504.08040}{{\tt arXiv:1504.08040}}].

\bibitem{Donnelly:2014fua}
W.~Donnelly and A.~C. Wall, {\it {Entanglement entropy of electromagnetic edge
  modes}},  {\em Phys. Rev. Lett.} {\bf 114} (2015), no.~11 111603,
  [\href{http://arxiv.org/abs/1412.1895}{{\tt arXiv:1412.1895}}].

\bibitem{Harlow:2015lma}
D.~Harlow, {\it {Wormholes, Emergent Gauge Fields, and the Weak Gravity
  Conjecture}},  {\em JHEP} {\bf 01} (2016) 122,
  [\href{http://arxiv.org/abs/1510.07911}{{\tt arXiv:1510.07911}}].

\bibitem{Donnelly:2016auv}
W.~Donnelly and L.~Freidel, {\it {Local subsystems in gauge theory and
  gravity}},  {\em JHEP} {\bf 09} (2016) 102,
  [\href{http://arxiv.org/abs/1601.04744}{{\tt arXiv:1601.04744}}].

\bibitem{Speranza:2017gxd}
A.~J. Speranza, {\it {Local phase space and edge modes for
  diffeomorphism-invariant theories}},  {\em JHEP} {\bf 02} (2018) 021,
  [\href{http://arxiv.org/abs/1706.05061}{{\tt arXiv:1706.05061}}].

\bibitem{Kirklin:2019xug}
J.~Kirklin, {\it {Unambiguous Phase Spaces for Subregions}},  {\em JHEP} {\bf
  03} (2019) 116, [\href{http://arxiv.org/abs/1901.09857}{{\tt
  arXiv:1901.09857}}].

\bibitem{DeWitt:1967yk}
B.~S. DeWitt, {\it {Quantum Theory of Gravity. 1. The Canonical Theory}},  {\em
  Phys. Rev.} {\bf 160} (1967) 1113--1148. [3,93(1987)].

\bibitem{Hayward:1993my}
G.~Hayward, {\it {Gravitational action for space-times with nonsmooth
  boundaries}},  {\em Phys. Rev. D} {\bf 47} (1993) 3275--3280.

\end{thebibliography}\endgroup
\end{document}